\documentclass[aps,floats]{revtex4}
\usepackage{amsmath}
\usepackage{graphicx,epsfig}

\begin{document}
\bibliographystyle {plain}

\def\oppropto{\mathop{\propto}} 
\def\opsimeq{\mathop{\simeq}}
\def\opoverderline{\mathop{\overline}}
\def\operarrow{\mathop{\longrightarrow}}
\def\opsim{\mathop{\sim}}

\def\fig#1#2{\includegraphics[height=#1]{#2}}
\def\figx#1#2{\includegraphics[width=#1]{#2}}


\title{Anomalous diffusion, Localization, Aging and Sub-aging effects \\
in trap models at very low temperature } 

\author{C\'ecile Monthus}
\affiliation{Service de Physique Th\'eorique, 
Unit\'e de recherche associ\'ee au CNRS, \\
DSM/CEA Saclay, 91191 Gif-sur-Yvette, France}

\begin{abstract}

We study in details the dynamics of the one dimensional symmetric trap model, via a real-space renormalization procedure which becomes exact in the limit of zero temperature. In this limit, the diffusion front in each sample consists in two delta peaks, which are completely out of equilibrium with each other. The statistics of the positions and weights of these delta peaks over the samples allows to obtain explicit results for all observables in the limit $T \to 0$. We first compute disorder averages of one-time observables, such as the diffusion front, the thermal width, the localization parameters, the two-particle correlation function, and the generating function of thermal cumulants of the position. We then study aging and sub-aging effects : our approach reproduces very simply the two different aging exponents and yields explicit forms for scaling functions of the various two-time correlations. We also extend the RSRG method to include systematic corrections to the previous zero temperature procedure via a series expansion in $T$. We then consider the generalized trap model with parameter $\alpha \in [0,1]$ and obtain that the large scale effective model at low temperature does not depend on $\alpha$ in any dimension, so that the only observables sensitive to $\alpha$ are those that measure the `local persistence', such as the probability to remain exactly in the same trap during a time interval. Finally, we extend our approach at a scaling level for the trap model in $d=2$ and obtain the two relevant time scales for aging properties.   

\end{abstract}

\maketitle

\section{Introduction}

\subsection{ Trap models as toy models to study aging phenomena}

Trap models provide a simple mechanism for aging \cite{feigelman,jp92}.
The basic phenomelogical idea is that the slow dynamics of glassy systems
is governed by metastable states defined as ``traps'' in the 
coarse-grained configurational complicated landscape. 
The distribution of the energy of the traps is usually taken to be
exponential
\begin{eqnarray}
\rho(E)= \theta(E) \frac{1}{T_g} e^{- \frac{E}{T_g}} 
\label{rhoe} 
\end{eqnarray}
On one hand, this exponential form describes
the lowest energies
in the Random Energy Model \cite{rem} and the distribution
of free energy of states 
in the replica theory of spin-glasses \cite{replica}.
On the other hand, it appears for the largest barriers  
 in the biased one-dimensional Sinai diffusion
\cite{feigelman,us_sinai} as well as in more complex disordered systems
such as fractals and percolation clusters \cite{rammal},
elastic manifolds \cite{vinokuretal}, bubble dynamics
in DNA \cite{bubbledna} and sequence alignment algorithms
 \cite{yuhwa}. The ubiquity of this exponential form actually comes \cite{jpmezard}
from the exponential tail 
of the Gumbel distribution which represents
one universality classes of extreme-value statistics \cite{extremebooks}.

The exponential density of energy (\ref{rhoe}) corresponds
for the Arrhenius trapping time $\tau = e^{\beta E}$
to the algebraic law
\begin{eqnarray}
q(\tau)
= \theta(\tau>1)  \frac{\mu}{\tau^{1+\mu}} 
 \label{qtau}
\end{eqnarray}
with the temperature-dependent exponent 
\begin{eqnarray}
\mu = \frac{T}{T_g}
\label{defmu}
\end{eqnarray}
At low temperatures $T<T_g$, the mean trapping time $\int d \tau 
\tau q(\tau)$ is infinite 
and this directly leads to aging effects. 
This mechanism shows that the presence of broad distribution 
of trapping times (\ref{qtau}) is rather generic at low temperatures,
since it simply emerges from the exponential tail of
extreme-value statistics for the energy barriers.

\subsection{ Previous results on aging properties in trap models }

The dynamics of the trap model has been studied in 
details in its the mean field version 
\cite{jpdean,cmjp,fielding,bertinjptg}
as well as in 
the one-dimensional {\it directed} version, in
relation with the biased Sinai diffusion 
\cite{feigelman,annphys,aslangul,laloux,comptejpb,c_directed}.
In both cases, aging properties are characterized by
scaling functions of the ratio $(t/t_w)$ of the two times involved.

More recently, it has been proposed in \cite{rinnmaassjp}
to study trap models on a hypercubic lattice in arbitrary dimension $d$
with the following generalized dynamics : the particle can jump from
site $i$ to any of the $2d$ nearest-neighbor sites $j$ with 
a hopping rate per unit time given by
\begin{eqnarray}
w_{i \to j} (\alpha)
= \frac{1}{2 d}  e^{+ \beta \alpha E_j - \beta(1-\alpha ) E_i }
\label{ratew}
\end{eqnarray}
in terms of the parameter $\alpha \in [0,1]$. 
The case $\alpha=0$ represents the usual trap model where the 
rate 
\begin{eqnarray}
w_{i \to j} (\alpha=0)
= \frac{1}{2 d \tau_i}  
\end{eqnarray}
depends 
only of the initial site via the trapping time $\tau_i=e^{\beta E_i}$ 
distributed with (\ref{qtau}) :
the particle spends at site $i$ a time $t_i$ distributed with the
exponential distribution of mean $\tau_i$
\begin{eqnarray}
f_{\tau_i}(t_i) = \frac{1}{\tau_i} e^{- \displaystyle \frac{t_i}{\tau_i} }
\label{ftau}
  \end{eqnarray}
and then jumps with equal probability $1/(2d)$ to one of its $(2d)$
nearest-neighbor sites.
Another interesting case is $\alpha= \frac{1}{2}$
where the rate $w_{i \to j}$ depend on the energy difference $(E_j-E_i)$.
 
Monte Carlo simulations and scaling arguments
 \cite{rinnmaassjp} have 
shown the possibility of a
 so-called ``sub-aging" behavior for the probability
$\Pi(t+t_w,t_w)$ of no jump during the interval $[t_w,t_w+t]$
\begin{eqnarray}
\Pi(t+t_w,t_w) \opsimeq   \Pi \left(  \frac{t}{t_w^{\nu(\alpha)}} \right)
\label{deffpi}
  \end{eqnarray}
with an exponent $\nu <1$ given in one dimension by \cite{rinnmaassjp} 
\begin{eqnarray}
\nu (\alpha) = \frac{1-\alpha}{1+\mu}
\end{eqnarray}
This exponent was proven to be exact by mathematicians, first for the
usual trap model $\alpha=0$ \cite{isopi}, and then for arbitrary $\alpha$
\cite{benarous}. On the other hand, 
the correlation function
 $C(t+t_w,t_w)$, defined as the probability 
to be at $(t+t_w)$ in the same trap it was as time $t_w$, was shown
to present a ``full aging" behavior
\begin{eqnarray}
C(t+t_w,t_w) \opsimeq {\cal C}_{\mu} \left(  \frac{t}{t_w} \right)
\label{deffc}
  \end{eqnarray}
 for the
usual trap model $\alpha=0$ \cite{isopi}, and then for arbitrary $\alpha$
\cite{benarous}.
So there are two different time scales $t_w^{\nu}$ and $t_w$
which play a role in the aging of this model.
Asymptotic forms have also been heuristically proposed 
and numerically tested in \cite{bertinjp}
for $\Pi(t+t_w,t_w)$ 
and $C(t+t_w,t_w)$. Finally, let us mention a recent interesting application : 
these properties of aging and subaging for the trap model are relevant 
to explain the numerical simulations
on the dynamics of denaturation bubbles in random DNA sequences
\cite{bubbledna}.  

\subsection{ Previous results on anomalous diffusion and localization properties} 

Apart from aging properties discussed above, trap models 
are interesting for their anomalous diffusion and localization properties.
In particular, in dimension $d=1$, the averaged diffusion front
is expected to take the following scaling form at large times
\cite{alexander,machta,bertinjp}
\begin{eqnarray}
\overline{ P(n,t \vert 0,0) }  \opsimeq_{t \to \infty}
\frac{1}{\xi(t)} g_{\mu} \left( \frac{n}{\xi(t)} \right)
\label{defg}
\end{eqnarray}
where the characteristic length
scale $\xi(t)$ follows the sub-diffusive behavior
\begin{eqnarray}
\xi(t)  \sim t^{\frac{\mu}{1+\mu}} 
\label{exponentspace} 
\end{eqnarray}
This exponent can be found via a simple scaling argument
on L\'evy sums \cite{alexander,argumentlevy}
or by a real space block-RG analysis \cite{machta}.
However, the scaling function itself $g(X)$ is not known, but
has recently been studied numerically in \cite{bertinjp}
together with asymptotic behaviors 
proposed in the limit $\mu \to 1$.

Another important issue concerns the localization properties.
The localization parameters 
\begin{eqnarray}
Y_k (t) = \sum_{n=-\infty}^{+\infty} 
 \overline{ P^k(n,t \vert 0,0)  } 
\label{defyk} 
\end{eqnarray}
represent the disorder averaged probabilities that $k$ independent particles
starting at site $0$ at time $0$  in the same random environment
are at the same site at time $t$.
It has been proven in \cite{isopi} that the limit $Y_2(\infty)$
is strictly positive in the full domain $0 \leq \mu <1$. 
The values of the limits $Y_k(\infty)$ have been numerically studied in 
\cite{bertinjp} with various approximations.

\subsection{ Summary of main results }

\label{mainres}

The aim of this article is to provide a probabilistic description,
sample by sample, of the symmetric trap model
in the limit of very low temperature $\mu \to 0$.
We have previously developed a similar analysis for the {\it directed}
version of the trap model \cite{c_directed}.
Here in the undirected version, each site may be visited many times
and this leads to essential changes.
 In particular, at time $t$, the important traps are
the traps having a trapping time $\tau>R(t)$,
where the scale $R(t)$ is not linear in $t$
as in the {\it directed}
version \cite{c_directed}, but is sub-linear in time
\begin{eqnarray}
  R(t) =  \left( \frac{t}{ {\tilde T}_0(\mu) } \right)^{\frac{1}{1+\mu}}
\end{eqnarray} 
where ${\tilde T}_0(\mu)$ may be expanded in $\mu$ as 
\begin{eqnarray}
   {\tilde T}_0(\mu)  
\opsimeq_{\mu \to 0} 2 e^{-1-\gamma_E}
\left[ 1  +O(\mu) \right]
\end{eqnarray} 
 The corresponding mean distance between these
important traps is then given by
 \begin{eqnarray}
 \xi(t)   
 = \xi_0(\mu) \ t^{\frac{\mu}{1+\mu}}  
\end{eqnarray}
where the exponent agrees with previous studies 
(\ref{exponentspace}) described above, and where 
the prefactor reads
 \begin{eqnarray}
 \xi_0(\mu) =1+O(\mu)  
\end{eqnarray}

In terms of these scales, we 
obtain the following explicit results in the limit $\mu \to 0$ : 

\begin{itemize}

\item Scaling function of the disorder averaged diffusion front (\ref{defg})
\begin{eqnarray}
g_{\mu}(X) = e^{- \vert X \vert}  \int_0^{+\infty} du
e^{-u} \frac{u}{\vert X \vert +u} +O(\mu) 
\end{eqnarray}

\item Localization parameters (\ref{defyk})
\begin{eqnarray}
Y_k (\mu) = 
   \frac{2}{(k+1)} +O(\mu)
\label{resyksummary}
\end{eqnarray}

\item Generating function of thermal cumulants 
\begin{eqnarray}
Z_{\mu}(s) \equiv \overline{ \ln < e^{-s \frac{n}{\xi(t)} }>}
 = \int_0^{+\infty} d \lambda e^{- \lambda} \lambda 
\left( \frac{s \lambda}{2} \coth \frac{s \lambda}{2}-1 \right)
+O(\mu)
 \end{eqnarray}
The series expansion in $s$
yields the disorder averages of rescaled 
thermal cumulants : the first ones are the thermal width 
\begin{eqnarray}
c_2(\mu)  \equiv \lim_{t \to \infty} \overline{\frac{<n^2>-<n>^2}
{\xi^2(t)} } =   1+O(\mu)
\end{eqnarray}
and the fourth cumulant
\begin{eqnarray}
c_4(\mu)  \equiv \lim_{t \to \infty} \overline{\frac{<n^4>-4<n^3><n>-3<n^2>^2+12<n^2><n>^2-6<n>^4}{\xi^4(t)}} 
= -4 +O(\mu)
\label{c4trap}
\end{eqnarray}
and more generally
\begin{eqnarray}
c_k(\mu)  
= (2k+1)! B_{2 k} +O(\mu)
\label{cktrap}
\end{eqnarray}
in terms of the Bernoulli numbers $B_{n}$.

\item Two-particle correlation function
\begin{eqnarray}
C(l,t) && \equiv \overline{  \sum_{n=0}^{+\infty} \sum_{m=0}^{+\infty}
P(n,t\vert 0,0) P(m,t\vert 0,0) \delta_{l,\vert n-m \vert} }  \opsimeq_{t \to \infty} Y_2(\mu) \delta_{l,0}
+ \frac{1}{ \xi(t) } {\cal C}_{\mu} \left( \frac{l}{\xi(t)} \right)
\label{correform}
 \end{eqnarray}
where the weight of the delta peak at the origin corresponds
as it should to the localization parameter $Y_2
=2/3 +O(\mu)$ (\ref{resyksummary}),
whereas the second part presents a scaling form of the variable $\lambda=\frac{l}{\xi(t)}$. The scaling function ${\cal C}_{\mu}$
reads
\begin{eqnarray}
 {\cal C}_{\mu} (\lambda)= e^{-\lambda} \frac{\lambda}{3}   +O(\mu) 
\label{correlong} 
\end{eqnarray}

\item  The probability $\Pi(t+t_w,t_w)$ of no jump during the interval $[t_w,t_w+t]$ takes the scaling form (\ref{deffpi})
\begin{eqnarray}
\Pi(t+t_w,t_w) = {\tilde \Pi}_{\mu} \left( g= \frac{t}{ R(t_w) } \right)
= {\tilde \Pi}_{\mu} \left( g= [ {\tilde T}_0(\mu) ]^{\frac{1}{1+\mu}}
 \frac{ t }{ t_w^{\frac{1}{1+\mu}}  } \right)
\end{eqnarray}
with the scaling function
 \begin{eqnarray}
 {\tilde \Pi}_{\mu}^{(0)} (g)
= \int_0^1 dz \mu z^{\mu-1}
 e^{- z  g }
\label{resumepi0}
\end{eqnarray}
In particular, we obtain the asymptotic behavior
\begin{eqnarray}
\Pi(t+t_w,t_w) \opsimeq_{\frac{ t }{ t_w^{\frac{1}{1+\mu}}  } \to + \infty} \left( \frac{t}{ t_w^{\frac{1}{1+\mu}}  } \right)^{-\mu}
\left[ \mu +O(\mu^2) \right]
\end{eqnarray}

 \item The probability $C(t+t_w,t_w)$ to be at time $(t+t_w)$ in the same trap
 at it was at time $t_w$ takes the scaling form (\ref{deffc})
\begin{eqnarray}
C(t+t_w,t_w) = {\tilde C}_{\mu} \left( h= \frac{t}{R^{1+\mu}(t_w)} \right)
= {\tilde C}_{\mu} \left( h= {\tilde T}_0(\mu) \frac{t}{t_w} \right)
\end{eqnarray}
with the scaling function which reads at lowest order in $\mu$ 
\begin{eqnarray}
 {\tilde C}_{\mu}^{(0)} (h)
=  {\tilde C}_{\mu} (h)
= \frac{2 \mu}{ ( 2  h )^{ \mu} }
 \int_0^{\sqrt{2  h} }  dz z^{1+2 \mu}
 K_1^2( z )
\end{eqnarray}
In particular, we obtain the asymptotic behavior
\begin{eqnarray}
C(t+t_w,t_w) \opsimeq_{ \frac{t}{t_w} \to \infty} 
\left( \frac{t}{t_w} \right)^{-\mu} \left[ \mu +O(\mu^2) \right]
\end{eqnarray}

\end{itemize}

We also extend the RSRG method to include systematic corrections
to the zero temperature procedure
via a series expansion in $\mu$ : 
the corrections of order $\mu$ of the observables
described above are given in Appendix \ref{correcmu1}.  
We also extend our analysis to
 the generalized trap models (\ref{ratew}) and obtain that 
the only observable that depends on $\alpha$ is
the two time correlation $\Pi(t,t')$, which takes the scaling form
\begin{eqnarray}
\Pi^{(\alpha)}_{\mu} (t+t_w,t) = {\tilde \Pi}_{\frac{\mu}{1-\alpha} }^{(0)} \left( v= \frac{t}{[R(t_w)]^{1-\alpha}} \right)
\end{eqnarray}
in terms of the result (\ref{resumepi0}) for the $\alpha=0$ case,
 but that otherwise all
other observables described above are exactly the same as
in the case $\alpha=0$. The reason is that the influence of
$\alpha$ is purely local around a renormalized trap
and does not change the renormalized effective model at large scales.
Finally, we also extend our RSRG approach to the trap model in dimension $d=2$
at a scaling level.

\subsection{ Organization of the paper }

The paper is organized as follows :

In Section \ref{rglandscape}, we defined the renormalized landscape
for the usual trap model $\alpha=0$
and study its properties : in particular, we obtain the relevant
length scale and the two relevant
time scales. In Section \ref{effectivemu0}, we describe
the effective dynamics in the limit $\mu \to 0$ and
compute one-time and two-times observables in this limit.
In Section \ref{correcfinitemu}, we study the corrections to the effective
dynamics at first order in $\mu$ and we describe the hierarchical structure
of the important traps that play a role at order $\mu^n$.
In Section \ref{secalpha}, we extend our approach to the generalized trap 
with parameter $\alpha \in [0,1]$. 
Section \ref{dim2} we extend our RSRG approach to the trap model
in $d=2$ at a scaling level. The conclusions are given in
Section \ref{conclusion},
and the Appendices contain more technical details.

\section{Definition and properties of the renormalized landscape}

\label{rglandscape}

\subsection{Notion of renormalized landscape at a scale $R$}

We wish to adapt the Real Space Renormalization procedure 
already defined for the Sinai model \cite{us_sinai} and for the directed
trap model \cite{c_directed} to the undirected trap model.
The basic idea is that the dynamics at large time is dominated by the
statistical properties of the large trapping times.
The renormalized landscape at scale $R$
is defined as follows : all traps with trapping time $\tau_i<R$
are decimated and replaced by a ``flat landscape'', whereas 
all traps with waiting time $\tau_i>R$ remain unchanged.
The distribution of the distance $l$ between two traps
of the renormalized landscape at scale $R$ reads
\begin{eqnarray}
  P_R (l) = \left[ 1- \int_R^{+\infty} d\tau q(\tau) \right]^{l-1}
 \int_R^{+\infty} d\tau q(\tau)
\end{eqnarray}
where the first part $[..]^{l-1}$ represents the probability that 
$(l-1)$ traps have a trapping time $\tau_i<R$, and where the last part
represents the probability that the $l^{\rm th}$ trap has a trapping time 
$\tau_i>R$. So the appropriate rescaled length variable
 at large scale $R$ is  
\begin{eqnarray}
\lambda= \frac{l }{R^{\mu} }
\label{length}
\end{eqnarray}
and the scaling distribution is simply exponential
\begin{eqnarray}
{\cal P} (\lambda)= e^{- \lambda}
\label{plambda}
\end{eqnarray}
The distribution of the trapping times of 
the traps in the renormalized landscape at scale $R$ 
is simply
\begin{eqnarray}
q_R(\tau)= \theta(\tau>R) \frac{q(\tau)}{ \int_R^{+\infty} d\tau' q(\tau')}
 = \theta(\tau>R) \frac{\mu}{\tau} \left( \frac{R}{\tau} \right)^{\mu}
 \label{qrtau}
\end{eqnarray}

In the directed version of the model, the particle visits each site only once,
and the RSRG analysis directly deals with the trapping times $\tau_i$.
However here in the undirected version of the model,
each site may be visited many times, and thus it is necessary
to introduce the notion of `escape time' as we now explain.

\subsection{Notion of `escape time' from a renormalized trap 
to another renormalized trap }

\begin{figure}

\centerline{\includegraphics[height=8cm]{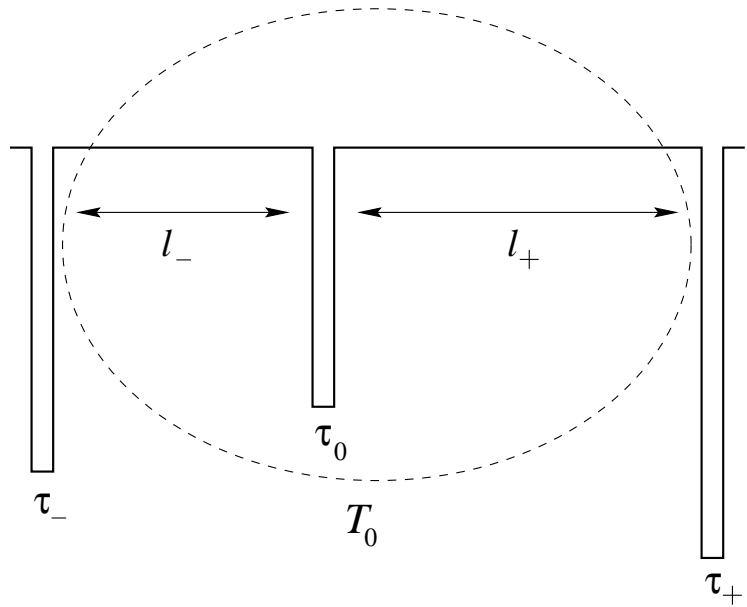}} 
\caption{ Definition of the escape time from a trap in the renormalized 
landscape :
the trap of escape time $\tau_0$ existing 
in the renormalized landscape at scale $R$
is surrounded by two
renormalized traps that are at distances $l_+$
and $l_-$ on each side. The escape time $T_0$ is
the mean time needed to reach either $\tau_+$
or $\tau_-$ when starting at $\tau_0$.   } 
\label{defescape}
\end{figure}

We now introduce the notion of the `escape time' $T$
from a trap $\tau_0$ existing in the renormalized landscape at scale $R$. 
This trap is surrounded by two
renormalized traps that are at distances $l_+$
and $l_-$ on each side (see Figure \ref{defescape}).
Whenever the particle escapes from the trap $\tau_0$, 
it can escape on either side with probability $(1/2)$.
If it escapes on the left, it will succeed to reach the trap $\tau_-$
with probability $1/l_-$, and if it escapes on the right,
it will succeed to reach the trap $\tau_+$
with probability $1/l_+$. Otherwise, it will be 
re-absorbed again by the trap $\tau_0$.

 \subsubsection{ Number of sojourns in a renormalized trap before escape to a neighbor renormalized trap}

 As a consequence, the probability $E_R(n)$ to 
need $(1+n)$ successive sojourns in the trap $\tau_0$ before the particle
succeeds to escape either to the trap $\tau_-$ or to the trap $\tau_+$ reads   
\begin{eqnarray}
E_R(n) && = \left[1- \frac{1}{2} \left( \frac{1}{l_+} + \frac{1}{l_-} \right) \right]^{n}
\frac{1}{2} \left( \frac{1}{l_+} + \frac{1}{l_-} \right)
\label{distrin}
\end{eqnarray}
For large $R$, since we have
$l_{\pm}=R^{\mu} \lambda_{\pm}$  (\ref{length}),
the number $n$ follows the same scaling : the rescaled variable
\begin{eqnarray}
w= \frac{n}{R^{\mu}}
\label{defw} 
\end{eqnarray}
is distributed exponentially
\begin{eqnarray}
{\cal E}(w)= a(\lambda_+,\lambda_-) e^{- w a(\lambda_+,\lambda_-) } 
\label{laww}
\end{eqnarray}
where the coefficient  
\begin{eqnarray}
 a(\lambda_+,\lambda_-) = \frac{1}{2} \left( \frac{1}{\lambda_+} + \frac{1}{\lambda_-} \right)
\end{eqnarray}
depends on the two rescaled distances to the next traps.
Its distribution $k(a)$ over the samples has for Laplace transform
in terms of the Bessel function $K_1$ (\ref{inteleadingtoK})
\begin{eqnarray}
{\hat k}(s) \equiv \int_0^{+\infty} da e^{-sa} k ( a )= \left[ \int_0^{+\infty} d\lambda  
{\cal P}(\lambda) e^{-\frac{s}{2 \lambda} } \right]^2
= 2 s K_1^2( \sqrt{2s})
\label{laplaceK}
\end{eqnarray}
 
In particular, the non-analytic behavior at small $s$ (\ref{K1zpetit})
\begin{eqnarray}
{\hat k}(s) \opsimeq_{s \to 0} 1 - s \ln \frac{1}{s} +O(s)
\end{eqnarray}
corresponds to the following algebraic decay at large $a$
\begin{eqnarray}
k(a) \opsimeq_{a \to \infty} \frac{1}{a^2}
\label{kalarge}
\end{eqnarray}

The decay at large $s$ (\ref{Kzgrand})
\begin{eqnarray}
{\hat k}(s) \opsimeq_{s \to \infty} \pi \sqrt{\frac{s}{2}} e^{-2 \sqrt{2 s}}
\end{eqnarray}
corresponds to the following essential singularity at small $a$
\begin{eqnarray}
k(a) \opsimeq_{a \to 0} \frac{\sqrt{2 \pi} }{
 a^{\frac{5}{2}} } e^{- \frac{2}{a} }
\label{kasmall}
\end{eqnarray}

 \subsubsection{ Total time spent inside a renormalized trap before escape to a neighbor renormalized trap}

Let us now consider the probability distribution $P_{in}(t_{in})$
of the total time $t_{in}$ spent inside the trap $\tau_0$ before its escape.
It can be decomposed into the number $n$ of sojourns,
where $n$ is distributed with (\ref{distrin})
\begin{eqnarray}
t_{in} = \sum_{i=1}^{1+n} t_i 
\end{eqnarray}
where $t_i$ is the time spent during the sojourn $i$
in the trap $\tau_0$, so it is distributed with
the exponential distribution (\ref{ftau}) with $\tau_i=\tau_0$.
Actually, since $n$ is large in the large $R$ limit,
we have the central-limit theorem
\begin{eqnarray}
t_{in} \opsimeq_{n \to \infty} n <t_i> = n \tau_0
\end{eqnarray}
This explains the numerical observation \cite{bertinjp}
that the results are unchanged if the particle spends
a time exactly equal to $\tau_i$ at each visit to site $i$,
instead of a random time $t_i$ distributed with (\ref{ftau}).

Since the number $n$ is distributed with
(\ref{defw}, \ref{laww}),
we finally obtain that $t_{in}$ is also exponentially distributed 
\begin{eqnarray}
{\hat P}_{in}(t_{in}) 
\opsimeq_{R \to \infty}  \frac{1}{T_0} e^{- \displaystyle \frac{t_{in}}{T_0} }
\label{pin}
\end{eqnarray}
with the characteristic time
\begin{eqnarray}
T_0 = \frac{1}{a} R^{\mu} \tau_0
\end{eqnarray}

Since the smallest trapping times 
existing in the renormalized landscape at scale $R$ is $\tau_0=R$,
the time spent inside the trap $\tau_0$ before it succeeds
to escape scales as
 \begin{eqnarray}
t_{in} \opsim_{R \to \infty}  R^{1+\mu} 
\label{scaletin}
\end{eqnarray}

 \subsubsection{ Total time spent during the unsuccessful excursions before the escape}

Among the $n$ unsuccessful
excursions, there are $m$ excursions on the left
and $(n-m)$ excursions on the left, where $m$ is distributed with 
the binomial distribution $2^{-n} C_n^m$. 
Since $n$ and $m$ are large, we again have a central-limit theorem
 \begin{eqnarray}
t_{out} = \sum_{i=1}^m t_i^- +\sum_{j=1}^{n-m} t_j^+ \opsimeq
m <t_-> + (n-m) <t_+> 
\label{tout} 
\end{eqnarray}
where $<t_{\pm}>$ 
represents the mean time needed to return to $0$ when starting
at $1$ without touching the point $l_{\pm}$ 
in a flat landscape. The asymptotic behavior (see Appendix 
\ref{appexcursions} for more details)
 \begin{eqnarray}
<t_{\pm}>  \opsimeq_{l_{\pm} \to \infty} \frac{l_{\pm}}{3}    
\end{eqnarray}
gives that the scale of $t_{out}$ at large $R$ reads
 \begin{eqnarray}
t_{out} \sim n l \sim R^{2 \mu} 
\label{scaletout}
\end{eqnarray}
which is negligible with respect to $t_{in}$ (\ref{scaletin})
for $\mu<1$.

 \subsubsection{ Time spent during the successful excursion to escape}

We finally consider the diffusion time $t_{diff}$ of the successful escape
to the neighbor renormalized landscape.
The free diffusion 
over a length $l \sim R^{\mu}$ takes a time of order $l^2$
(see Appendix 
\ref{appexcursions} for more details) and thus
the scale of $t_{diff}$ at large $R$
 \begin{eqnarray}
t_{diff} \sim R^{2\mu} 
\label{tdiff}
\end{eqnarray}
is the same as $t_{out}$ (\ref{scaletout})
but is negligible with respect to $t_{in}$ (\ref{scaletin})
for $\mu<1$.

\subsubsection{ Conclusion}

So we obtain that the total time 
\begin{eqnarray}
t_{esc} = t_{in}+t_{out}+t_{diff}
\label{tesc3contri}
\end{eqnarray}
 needed to escape is actually simply given by
the time $t_{in}$ spent inside the trap $\tau_0$.
So the distribution of $t_{esc}$
is given by the exponential (\ref{pin})
with the escape time $T_0$.

In conclusion, a trap 
 of the renormalized landscape at scale $R$
has a trapping time $\tau$ distributed with (\ref{qrtau}), but has 
an `escape time' proportional to $\tau$ 
\begin{eqnarray}
T=\frac{1}{a} R^{\mu} \tau
\end{eqnarray}
with a factor $R^{\mu}$ that explains the 
occurrence of two different time scales in this model,
and with a prefactor $a$ distributed with (\ref{laplaceK})

\subsection{Distribution of `escape times' in the renormalized landscape}

The distribution of the escape time $T$ in the renormalized landscape
at scale $R$ reads
\begin{eqnarray}
Q_R(T) = \int_R^{+\infty} d\tau q_R(\tau) \int_0^{+\infty} da k(a)
\delta(T- \frac{1}{a} R^{\mu} \tau)
\end{eqnarray}
It thus presents the scaling form
\begin{eqnarray}
Q_R(T) = 
 \frac{1}{R^{1+\mu}} {\cal Q}_{\mu} \left(
{\tilde T} = \frac{T}{R^{1+\mu}} \right)
\label{scalingqt}
\end{eqnarray}
where the scaling function reads
\begin{eqnarray}
{\cal Q}_{\mu} \left( {\tilde T} \right) =  \frac{\mu}{{\tilde T}^{1+\mu} } 
\int_0^{+\infty} da \frac{k(a)}{a^{\mu} } \theta(a > \frac{1}{{\tilde T}} )
 \label{qrtfull}
\end{eqnarray}

In particular, for large ${\tilde T}$, there is the same algebraic decay
with index $(1+\mu)$ as for the distribution of trapping times
\begin{eqnarray}
{\cal Q}_{\mu} \left( {\tilde T} \right) 
\opsimeq_{{\tilde T} \to \infty} \frac{\mu}{{\tilde T}^{1+\mu} }
 c(\mu)
 \label{qrtasymp}
\end{eqnarray}
where the constant $c(\mu)$ may be computed from 
the Laplace transform (\ref{laplaceK}) using (\ref{inteK2power}) 
\begin{eqnarray}
c(\mu)  \equiv \int_{0}^{+\infty} da \frac{k(a)}{a^{\mu}}
 = \frac{1}{\Gamma(\mu) } \int_0^{+\infty} ds s^{\mu-1} 
{\tilde k} (s)
= 2^{\mu} \frac{(1+\mu) \Gamma^3(1+\mu) }{\Gamma(2+2 \mu)} 
 = 1-\mu (1+\gamma_E - \ln 2) +O(\mu^2)
\label{calcmu}
\end{eqnarray}

For small ${\tilde T}$, we
use the asymptotic behavior of $k(a)$ at large $a$ (\ref{kalarge})
\begin{eqnarray}
k(a) 
\opsimeq_{a \to 0} \frac{1}{a^2} 
+ \frac{\ln a+(\ln 2 -1 -\gamma_E)}{a^3} +O( \frac{\ln a}{a^4} )
\end{eqnarray}
to obtain 
\begin{eqnarray}
{\cal Q}_{\mu} \left( {\tilde T} \right) 
\opsimeq_{{\tilde T} \to 0} \frac{\mu}{1+\mu} 
+ \frac{\mu}{2+\mu} {\tilde T} \left( \ln \frac{1}{{\tilde T} }
+\ln 2 -\gamma_E -1 + \frac{1}{2+\mu}\right) + O \left({\tilde T}^2  \ln \frac{1}{{\tilde T} } \right)
\end{eqnarray}

Using (\ref{laplaceK}), the most convenient way
to characterize the scaling function in closed form
is by the following transform in terms of a variable $\omega$
\begin{eqnarray}
\int_0^{+\infty} d{\tilde T} e^{- \frac{\omega^2 }{ 2 {\tilde T}} }
{\cal Q}_{\mu} \left( {\tilde T} \right)  = 
\int_0^{1}  \mu dv v^{\mu-1}
\int_0^{+\infty} da k(a)
  e^{- \frac{\omega^2 }{ 2} v a }= \frac{2 \mu}{ \omega^{2 \mu} } \int_0^{\omega}  dz z^{1+2 \mu}
 K_1^2( z )
\label{qtildetransform}
\end{eqnarray}

\subsection{Choice of the renormalization scale $R$ as a function of time}

For small $\mu$, the probability distribution ${\cal Q}_{\mu} \left( {\tilde T} \right)$ is dominated by its long tail (\ref{qrtasymp}), and we may approximate
it by
\begin{eqnarray}
{\cal Q}_{\mu} \left( {\tilde T} \right) 
\simeq \theta( {\tilde T} > {\tilde T}_0 (\mu) )
\frac{\mu}{{\tilde T}} \left( \frac{{\tilde T}_0 (\mu)}
{{\tilde T}}\right)^{\mu} 
\end{eqnarray}
where the cut-off ${\tilde T}_0$ chosen to preserve the normalization
is determined by the coefficient for
the long tail part (\ref{qrtasymp},\ref{calcmu}) 
\begin{eqnarray}
 {\tilde T}_0(\mu) = \left( c(\mu)  \right)^{\frac{1}{\mu} } 
\opsimeq_{\mu \to 0} 2 e^{-1-\gamma_E}
\left[ 1 + \mu \frac{18-\pi^2}{12} +O(\mu^2) \right]
\label{tildet0mu0}
\end{eqnarray}
For the unrescaled probability distribution (\ref{scalingqt}),
this corresponds to the cut-off
\begin{eqnarray}
T_0( \mu)= R^{1+\mu} {\tilde T}_0(\mu)
\end{eqnarray}
 It is thus convenient to associate at time $t$ the renormalization scale $R(t)$
such that
\begin{eqnarray}
T_0( \mu) = t
 \end{eqnarray}
meaning that at time $t$, only traps with escape times $T>t$
have been kept, whereas all traps with escape times $T<t$
have been removed and replaced by a flat landscape.
This leads to the explicit choice 
\begin{eqnarray}
  R(t) =  
  \left( \frac{t}{ {\tilde T}_0(\mu) } \right)^{\frac{1}{1+\mu}}
\label{roft}
\end{eqnarray} 
 The corresponding mean distance between traps reads at this 
renormalization scale (\ref{length})
 \begin{eqnarray}
 \xi(t) \equiv [R(t)]^{\mu}  
 = \xi_0(\mu) t^{\frac{\mu}{1+\mu}}  
\label{xit}
\end{eqnarray}
with 
the prefactor 
 \begin{eqnarray}
 \xi_0(\mu) =\left(  c(\mu)  \right)^{-\frac{1}{1+\mu}}
= 1+\mu (1+\gamma_E - \ln 2) +O(\mu^2)  
\end{eqnarray}

\section{Effective model at large time in the limit $\mu \to 0$}

\label{effectivemu0}

\subsection{ Effective rules for the dynamics}  

The prescription for the dynamics is as follows :

At time $t$,  the particle starting at the origin $O$ 
 will be at time $t$ either in the first trap $M_+$ of the renormalized landscape at scale $R(t)$
 on its right or in the 
 first trap $M_-$ of the renormalized landscape
 on its left.
 The weight $p_{M_+}$ of the trap $M_+$ 
 is given by the probability to reach $M_+$ before $M_-$ 
 for a particle performing a pure random walk, so it is simply given by
 the ratio of the distances from its starting point
 \begin{eqnarray}
p_{[M_-M_+]}(M_+ \vert 0) && = \frac{ M_-O  }{ M_-M_+  } = \frac{l_-}{ l_++l_- }  \\
p_{[M_-M_+]}(M_- \vert 0) && = \frac{ OM_+  }{ M_-M_+  } = \frac{l_+}{ l_++l_- }
\label{rulespplusmoins}
\end{eqnarray}

This rule for the effective dynamics is consistent upon iteration.
Suppose there are three consecutive traps :
the trap $M_-$ is at a distance $l_-$ from the origin on the left,
the trap $M_+$ is at a distance $l_+$ from the origin on the right,
and the trap $M_{++}$ is at a distance $l$ from the trap $M_+$ on the right.
Suppose that the trap $M_+$ is decimated before the traps $M_-$ and $M_{++}$.
The new weights for the traps $M_-$ and $M_{++}$ become
 \begin{eqnarray}
p_{M_-}' && = p_{[M_-M_+]}(M_- \vert 0) + 
p_{[M_-M_{++}]} (M_- \vert M_+) p_{[M_-M_+]}(M_+ \vert 0) 
= \frac{ l_++l }{ l_-+l_+ +l } = p_{[M_-M_{++}]}(M_- \vert 0) \\
p_{M_{++}}' && =  p_{[M_-M_{++}]}(M_{++} \vert M_+) 
p_{[M_-M_+]}(M_+ \vert 0) 
= \frac{ l_-}{ l_- +l_+ +l }  = p_{[M_-M_{++}]}(M_{++} \vert 0)
\label{ruledecim}
\end{eqnarray}
and thus the rules (\ref{rulespplusmoins}) 
for the occupancies of renormalized traps
are consistent upon decimation of traps in the renormalized landscape.

\subsection{ Diffusion front}

In this effective model, 
the diffusion front in a given sample thus reads
\begin{eqnarray}
 P^{(0)}_t(n) = \frac{1}{\xi(t) } {\cal P}^{(0)} \left( X= \frac{n}{\xi(t) }  \right)
\end{eqnarray}
where the scaling function reads
\begin{eqnarray}
 {\cal P}^{(0)} \left( X  \right) = 
\frac{X_+}{X_++X_-} \delta(X+X_-)
+\frac{X_-}{X_++X_-} \delta(X-X_+)
\label{pMM}
\end{eqnarray}
in terms of the two rescaled distances $X_{\pm}$ between the origin and the 
nearest renormalized traps.
Since the joint distribution of the two
rescaled distances is completely factorized 
 \begin{eqnarray}
{\cal D}( X_+,X_-) = \theta(X_+) \theta(X_-) e^{-X_+ -X_-} 
\label{measure0}
\end{eqnarray}
 we obtain that the scaling function 
for the disorder averaged diffusion front (\ref{defg})
reads at lowest order in $\mu$
\begin{eqnarray}
g^{(0)}(X) && = \int_0^{+\infty} dX_+  \int_0^{+\infty} dX_- 
{\cal D}( X_+,X_-)
 {\cal P}^{(0)} ( X  ) = e^{- \vert X \vert}  \int_0^{+\infty} du
e^{-u} \frac{u}{\vert X \vert +u} 
\end{eqnarray}
In particular, its asymptotic behaviors read
\begin{eqnarray}
g^{(0)}(X) && \opsimeq_{ \vert X \vert\to \infty} 
 \frac{1}{\vert X \vert }  e^{- \vert X \vert} \\
g^{(0)}(X) && \opsimeq_{ \vert X \vert\to 0} 
1- \vert X \vert
 \left[ \ln \frac{1}{\vert X \vert } -\gamma_E \right] +O \left(
\vert X \vert^2
  \ln \frac{1}{\vert X \vert } \right)
\end{eqnarray}
It is interesting to compare with the simple exponential front $e^{-X}$
obtained in the {\it directed} version of the same trap model.
At infinity, the front is reduced by the power $\frac{1}{\vert X \vert }$
with respect to the exponential representing the distribution of the
distance to a renormalized trap, because of the probability $1/l$
to escape to this trap instead of being absorbed by
a nearer renormalized trap on the other side.
On the other hand, near the origin, the front is enhanced by the log
term because it is more probable to be absorbed by a
trap which happens to be very near instead of being absorbed by
the renormalized trap on the other side.

\subsection{ Localization parameters}

In a given sample, the localization parameters read at this order
\begin{eqnarray}
Y_k^{(0)}= p_{M_+}^k+p_{M_-}^k
\end{eqnarray}
and thus averaging over the samples yields
\begin{eqnarray}
\overline{Y_k^{(0)}} = 
\int_0^{+\infty} dX_+  \int_0^{+\infty} dX_-
{\cal D}( X_+,X_-) \frac{ X_+^k + X_-^k }{ (X_++X_-)^k}  
=   \frac{2}{(k+1)}
\end{eqnarray}
The agreement with the numerical simulations of \cite{bertinjp}
obtaining $Y_2 \to 2/3$ and $Y_3 \to 1/2$ is the clearest numerical evidence
of the validity of the RSRG effective dynamics with only two relevant
traps in each sample in the limit $\mu \to 0$.

\subsection{ Thermal width}

In a given sample, the thermal width reads at this order
\begin{eqnarray}
<\Delta n^2(t)>^{(0)} = (n_{M_+} +n_{n_-})^2 p_{M_+} p_{M_-} = n_{M_+} n_{M_-}
\end{eqnarray} 
and thus we obtain after averaging over the samples
\begin{eqnarray}
c_2(\mu \to 0) \equiv \lim_{t \to \infty} 
\frac{\overline{<\Delta n^2(t)>}}{\xi^2(t)} = 
\int_0^{+\infty} dX_+  \int_0^{+\infty} dX_-
 {\cal D}( X_+,X_-) X_+ X_-   = 1
\end{eqnarray} 

\subsection{ Disorder averages of thermal cumulants}

The generating function of disorder averages of
rescaled thermal cumulants
 reads
\begin{eqnarray}
Z(s) \equiv \overline{ \ln < e^{-s \frac{n}{\xi(t)} } >  }
 \end{eqnarray}
thus reads at lowest order
\begin{eqnarray}
Z^{(0)}(s) && = \int_0^{+\infty} dX_+  \int_0^{+\infty} dX_-
 {\cal D}( X_+,X_-)  \ln \left( \frac{X_+}{X_++X_-} e^{s X_-}
+\frac{X_-}{X_++X_-} e^{-s X_+} \right) 
\\
&& = \int_0^{+\infty} d \lambda e^{- \lambda} \lambda 
\left( \frac{s \lambda}{2} \coth \frac{s \lambda}{2}-1 \right)
 \end{eqnarray}
The series expansion in $s$
gives the disorder averages of
the rescaled thermal cumulants (\ref{cktrap}).

\subsection{ Two-particle correlation function }

The two-particle correlation function reads
\begin{eqnarray}
 C(l,t) \equiv \sum_{n=0}^{+\infty} \sum_{m=0}^{+\infty}
\overline{P(n) P(m) }\delta_{l,\vert n-m \vert} 
\opsimeq_{t \to \infty} Y_2^{(0)} \delta_{l,0}
+ \frac{1}{\xi(t) } {\cal C}_{\mu} \left( \lambda= \frac{l}{\xi(t)} \right)
\end{eqnarray} 
where the weight of the delta peak corresponds as it should to the localization parameter $ Y_2^{(0)} = \frac{2}{3}$
discussed above, whereas the scaling function of the long-ranged part reads at lowest order
\begin{eqnarray}
  {\cal C}_{\mu}^{(0)} \left( \lambda \right) =
 \int_0^{+\infty} dX_+  \int_0^{+\infty} dX_-
{\cal D}( X_+,X_-) \frac{2 X_+ X_-}{(X_++X_-)^2} \delta_{\lambda,X_++X_-} 
 = e^{-\lambda}   \frac{\lambda}{3}  
\end{eqnarray}

\subsection{Aging and Sub-aging properties}

As explained in the introduction, there are two different
correlation functions which present different aging properties
\cite{isopi,bertinjp,benarous}. 
We now very simply recover within our framework
the expected sub-aging and aging exponents,
 and moreover compute the scaling functions
in the limit $T \to 0$.

\subsubsection{ Probability $\Pi(t+t_w,t_w)$ of no jump during he interval $[t_w,t_w+t]$}

The probability $\Pi(t+t_w,t_w)$ of no jump during he interval $[t_w,t_w+t]$
is directly related to the probability $\psi_{t_w}(\tau)$
 to be at time $t_w$ in a trap of trapping time $\tau$
via the integral
\begin{eqnarray}
\Pi(t+t_w,t_w) = \int_0^{+\infty} d\tau \psi_{t_w}(\tau) e^{- \frac{t}{\tau} }
\label{defpittprime}
\end{eqnarray}

In our approach, we have assumed as a starting
point that at lowest order in $\mu$, the probability
$\psi_t(\tau)$ was given by the trapping time distribution in the renormalized
landscape at scale $R(t)$ (\ref{roft})
\begin{eqnarray}
 \psi_{t}(\tau) = q_{R(t)}(\tau) = \theta(\tau>R(t)) \mu \frac{ R^{\mu}(t)}
{\tau^{1+\mu}}
\end{eqnarray}

So we obtain the scaling form (\ref{deffpi})
\begin{eqnarray}
\Pi(t+t_w,t_w) = {\tilde \Pi}_{\mu} \left( g= \frac{t}{ R(t_w) } \right)
= {\tilde \Pi}_{\mu} \left( g= [ {\tilde T}_0(\mu) ]^{\frac{1}{1+\mu}} \frac{ t }{ t_w^{\frac{1}{1+\mu}}  } \right)
\label{piscalingform}
\end{eqnarray}
and the scaling function reads at lowest order in $\mu$
\begin{eqnarray}
 {\tilde \Pi}_{\mu}^{(0)} (g)
= \int_0^1 dz \mu z^{\mu-1}
 e^{- z  g }
\label{piscal0}
\end{eqnarray}

In particular, its asymptotic behaviors reads
\begin{eqnarray}
 {\tilde \Pi}_{\mu}^{(0)} (g)
&& \opsimeq_{g \to 0} 1 -g  \frac{\mu}{1+\mu}
  +O(g^2) 
\\
 {\tilde \Pi}_{\mu}^{(0)} (g)
&&  \opsimeq_{g \to + \infty} \frac{\Gamma(1+\mu)}{g^{\mu} }
\end{eqnarray}

\subsubsection{ 
 Probability $C(t+t_w,t_w)$ to be at time $(t+t_w)$ in the same trap
 at it was at time $t_w$ }

According to the analysis of the escape time 
from a renormalized trap to a next nearest one,
the total escape time (\ref{tesc3contri}) is actually
dominated by the time $t_{in}$ spent inside the renormalized trap.
As a consequence, we have
\begin{eqnarray}
C(t+t_w,t_w) = \int dT \phi_{t_w}(T) e^{- \frac{t}{T} }
\label{ctophi}
\end{eqnarray}
where $\phi_{t_w}(T)$ represents the probability to be 
in a trap of escape time $T$ in the renormalized landscape
at scale $R(t_w)$. 

At lowest order in $\mu$, $\phi_t(T)$ is simply given by
the probability $Q_{R(t)}(T)$ given in 
(\ref{scalingqt}).
So we obtain the scaling form
\begin{eqnarray}
C(t+t_w,t_w) = {\tilde C}_{\mu} \left( h= \frac{t}{R^{1+\mu}(t_w)} \right)
= {\tilde C}_{\mu} \left( h= {\tilde T}_0(\mu) \frac{t}{t_w} \right)
\label{defcscalw}
\end{eqnarray}
where the scaling function reads at lowest order in $\mu$
 \begin{eqnarray}
 {\tilde C}_{\mu} (h) = \int_0^{+\infty} d {\tilde T} {\cal Q}_{\mu} ( {\tilde T} ) e^{-   \frac{h}{\tilde T}}
\end{eqnarray}
in terms of the function ${\cal Q}_{\mu} ( {\tilde T} )$
introduced in (\ref{qrtfull}).
Using (\ref{qtildetransform}), we thus obtain the scaling function
in terms of the Bessel function $K_1$ as

\begin{eqnarray}
 {\tilde C}_{\mu} (h)
= \frac{2 \mu}{ ( 2  h )^{ \mu} }
 \int_0^{\sqrt{2  h} }  dz z^{1+2 \mu}
 K_1^2( z )
\label{correscaling}
\end{eqnarray}

In particular, the asymptotic behaviors read
\begin{eqnarray}
 {\tilde C}_{\mu} (h) && \opsimeq_{ h \to 0} 
1 - \frac{\mu  }{1+\mu} h \ln \frac{1}{h} +O(h)
\\ {\tilde C}_{\mu} (h) && \opsimeq_{ h \to \infty} 
\frac{\Gamma(1+\mu)}{h^{\mu} }
\end{eqnarray}

\section{ Corrections to the zero temperature effective dynamics }

\label{correcfinitemu}

In the previous Section, to compute all observables in the limit $\mu \to 0$,
we have considered that the distribution of
the escape time was infinitely broad in the following sense :
 all traps with $T_i<t$ were such that $\frac{T_i}{t} \sim 0$, 
whereas all traps with $T_i>t$
were such that $\frac{T_i}{t} \sim +\infty$. 
For finite $\mu$, we have to take into account
 that these ratios are not really zero or infinite.
This can be done in a systematic procedure
order by order in $\mu$ as we now explain.

\subsection{New effects at order $\mu$ }

\label{dispersionmu}

\begin{figure}

\centerline{\includegraphics[height=8cm]{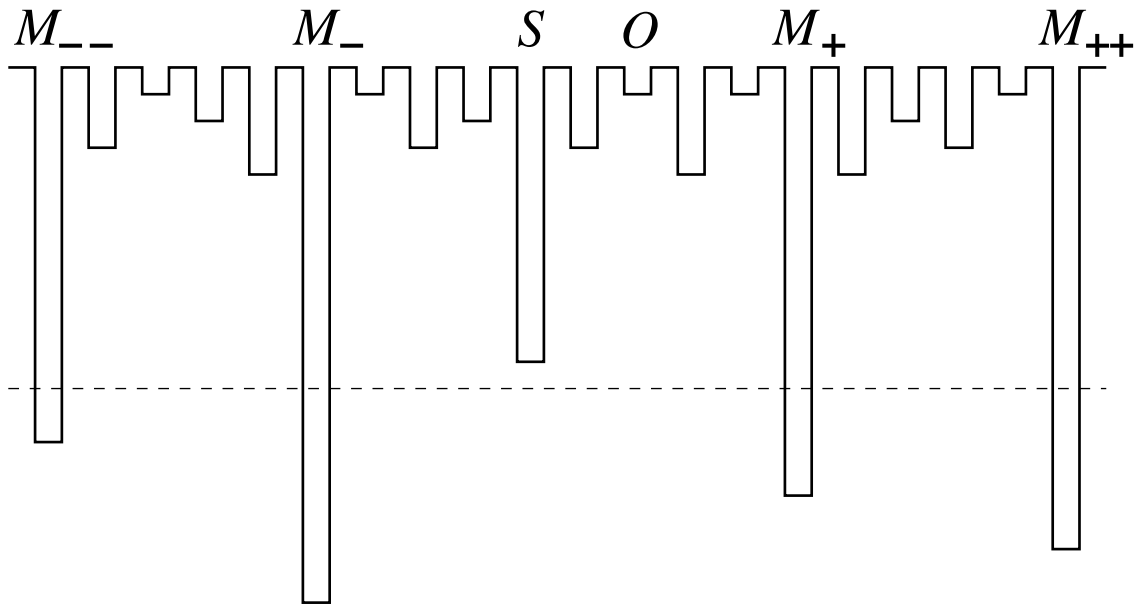}} 
\caption{ Construction of the important traps in a given sample sample
for a particle starting at the origin.
The dashed line separates the ``small" traps (that have a trapping time
smaller than $R(t)$) from the ``big" traps (that have a trapping time bigger 
than $R(t)$). The first big traps on each side called $M_+$ 
and $M_-$ are occupied with
a weight of order $O(\mu^0)$. The next big traps $M_{++}$, $M_{--}$
 and the biggest 
small trap $S$ in the interval $]M_-,M_+[$ are occupied with weights of order $O(\mu)$.
  } 
\label{trapsym}
\end{figure}

At first order in $\mu$, we need to consider the following effects 
(see Figure \ref{trapsym}) :

\begin{itemize}

\item
 A trap $M$ of the renormalized landscape has a escape time $T_{M}$ which is not infinite. 
 There is a small probability $(1-e^{-\frac{t}{T_{M}}})$ that the particle has already
  escaped from this trap at time $t$.
  
  If it has escaped, it has been absorbed by one of the two renormalized
 neighbors, with probabilities given by the ratios of the distances.
We will say that the particle is ``in advance" with respect
to the effective dynamics of the limit $\mu \to 0$.

\item
 The biggest trap in an interval between two renormalized traps
that we will call $S$ , 
has an escape time $T_{S}<t$ which is not zero and thus there is a small probability of order $e^{-\frac{t}{T_{S}}}$ that the particle
 is still trapped in $S$ at time $t$.
We will say that the particle is ``late" with respect
to the effective dynamics of the limit $\mu \to 0$.  

\end{itemize}

The first corrections of order $\mu$
for various observables are given 
in Appendix \ref{correcmu1}.
From a technical point of view, it turns out 
that the computation of averages
over configurations
becomes rapidly much more involved than the similar computations
for the directed case \cite{c_directed}. This is due to the fact
that the escape time depends on both the trapping time and
the distances to the neighbor traps. 
So here, contrary to the directed case
where we have computed corrections up to order $\mu^2$ \cite{c_directed},
we will not go further into the explicit corrections of higher orders,
but simply describe the hierarchical structure of the important traps
that appear at a given order in $\mu$.

\subsection {  Hierarchical structure of the important traps}

\label{hierarchie}

The procedure that we have described
up to order $\mu$ can be generalized 
at an arbitrary order $n$ as follows : all observables at order $\mu^n$
can be obtained by considering a dispersion of the thermal packet
over at most $(2+n)$ traps, that have to be chosen among a 
certain number
$\Omega_n$ of possible configurations of the traps. 
Our aim here is simply
 to get some insight into the set of important
traps that play a role at a given order $n$.

At order $\mu^n$, the important traps are :

\begin{itemize}

\item the main traps $M_-$ and $M_+$

\item the next $n$ large renormalized traps on each side :

$M_{2+}$,...$M_{(n+1)+}$ on the half line $]M_+,+\infty[$

$M_{(n+1)-}$,...$M_{2-}$ on the half line $]-\infty,M_-[$

\item the $n$ biggest traps $S_1^{(0)}$...$S_n^{(0)}$ among the small traps 
in the interval $]M_-,M_+[$

\item the $(n-1)$ biggest traps $S_2^{+(1)}$...$S_n^{+(1)}$ among the small traps
in the interval $]M_+,M_{2+}[$, and the $(n-1)$ biggest traps $S_2^{-(1)}$...$S_n^{-(1)}$ among the small traps
in the interval $]M_{2-},M_{-}[$

\item the $(n-2)$ biggest traps $S^{+(2)}_3$...$S^{+(2)}_n$ among the small traps in the interval $]M_{2+},M_{3+}[$, and the $(n-2)$ biggest traps $S^{-(2)}_3$...$S^{-(2)}_n$ among the small traps in the interval $]M_{3-},M_{2-}[$

\item...

\item the biggest trap 
$S^{+(n-1)}_n$ among the small traps in the interval
 $]M_{(n-1)+},M_{n+}[$, and the biggest trap 
$S^{-(n-1)}_n$ among the small traps in the interval
 $]M_{n-},M_{(n-1)-}[$

\end{itemize}

The total number of traps is thus  
\begin{eqnarray} 
T_n=2+n+2 \sum_{i=1}^n i = 2+n(n+2)
\label{numbertn}
\end{eqnarray}
which generalizes the number $T_0=2$ ($M_-$,$M_+$) at order $n=0$
and the number $T_1=5$ ($M_-$,$M_+$,$M_{--}$,$M_{++}$,$S$)
at order $n=1$

With these $T_n$ traps, one has to construct the possible 
$\Omega_n$ configurations of
$(2+n)$ traps, that are ordered in positions,
and that contribute
up to order $\mu^n$, as explained
in more details in Appendix \ref{configordern}.

\subsection { Discussion }

In conclusion, the effective dynamics for the trap model
valid in the limit $\mu \to 0$
is also a good starting point to study the full aging phase $0<\mu<1$,
since one can build a systematic series expansion in $\mu$
for all observables. Moreover, the hierarchical structure
of the traps that are important at order $n$
 give a clear insight into the structure of the dynamics
for finite $\mu$.

\section{ Properties of the renormalized
landscape for the generalized model}

\label{secalpha}

We now consider the generalized model 
defined by the hopping rates (\ref{ratew}) with parameter $\alpha \in [0,1]$
and study the changes with respect to the usual
trap model corresponding to the special case $\alpha=0$
that we have studied in the previous Sections.

\subsection{ Definition of the renormalized landscape}

Here the renormalized landscape is defined for the energy variable :
at scale $\Gamma$, all traps with energy $E< \Gamma$ are decimated
and replaced by a flat landscape, whereas all traps with energy
$E> \Gamma$ remain unchanged. 
So the distribution of the energy in the renormalized landscape
simply reads (\ref{rhoe})
\begin{eqnarray}
\rho_{\Gamma}(E)= \theta(E>\Gamma) \frac{1}{T_g} e^{- \frac{E-\Gamma}{T_g}} 
\end{eqnarray}

\subsection{ Trapping-time of a renormalized trap}

\begin{figure}

\centerline{\includegraphics[height=8cm]{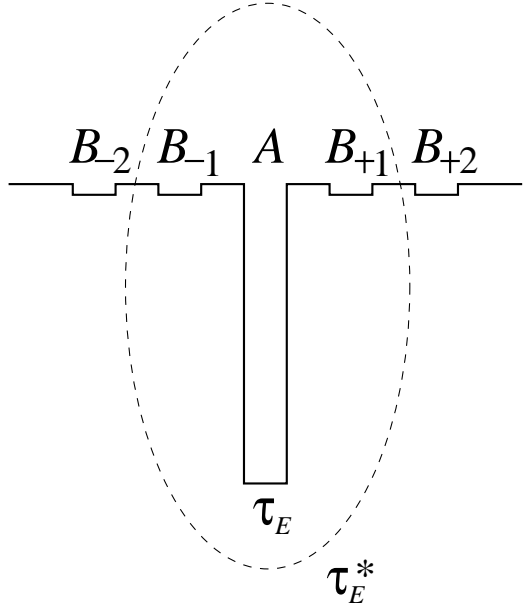}} 
\caption{ Definition of the two escape times :
a deep trap $A$ of the renormalized landscape 
has for nearest neighbors
$B_{\pm 1}$ and for next nearest neighbors
$B_{\pm 2}$. In the case $\alpha>0$, the trapping time $\tau_E$
of $A$ alone depends on $\alpha$, whereas
 the trapping time $\tau_E^*$ of the 
interval made with the three consecutive traps $B_{-1}AB_{+1}$ does not depend on $\alpha$ 
  } 
\label{BAB}
\end{figure}

Let us consider a trap called $A$ with energy
 $E> \Gamma$ existing in the renormalized landscape :
it is surrounded by traps $(B_1,B_2)$ on the right
and $(B_{-1},B_{-2})$ on the left which have $E \sim 0$ (see Figure \ref{BAB}).
As a consequence, the hopping rates to escape from $A$ are 
\begin{eqnarray}
w_{A \to B_{\pm 1}} = \frac{1}{2} e^{- \beta (1-\alpha) E}
\end{eqnarray}
so we may associate to the trap $A$ the trapping time
\begin{eqnarray}
\tau_E = e^{ \beta (1-\alpha) E}
\label{tauE}
\end{eqnarray}
The distribution of $\tau_E$ in the renormalized landscape
at scale $\Gamma$ reads
\begin{eqnarray}
q^{(\alpha)}_{\Gamma} (\tau) = \theta( \tau> \tau_{\Gamma} )
\frac{ \mu(\alpha) }{\tau} 
\left( \frac{\tau_{\Gamma} } {\tau} \right)^{\mu(\alpha)}
\label{qalpha}
\end{eqnarray}
with the cut-off 
\begin{eqnarray}
\tau_{\Gamma} =e^{ \beta (1-\alpha) \Gamma}
\label{taugamma}
\end{eqnarray}
 and with the new exponent 
\begin{eqnarray}
\mu(\alpha)=\frac{\mu}{1-\alpha}
\label{mualpha}
\end{eqnarray}
instead of (\ref{defmu}).

\subsection{ Trapping-time of the vicinity
of a renormalized trap}

The hopping rates from the site $B_1$ ( or $B_{-1}$) 
nearest neighbor of the renormalized trap $A$ of energy $E$ read
\begin{eqnarray}
w_{ B_{1} \to A} && = \frac{1}{2} e^{\beta \alpha E} \\
w_{ B_{1} \to B_2} && = \frac{1}{2}
\end{eqnarray}
so there is a very high probability to 
return immediately to the deep trap $A$.

We are interested into the probability distribution $P_*(t^*)$
of the total time $t^*$ spent at $A$
before the first passage at ${B_2} $ or $ B_{-2}$
when starting at $A$.
We may decompose $t^*$ into the number $k$ of sojourns
at $A$ needed  
\begin{eqnarray}
t^*=\sum_{i=1}^k t_i
\label{tetoile}
\end{eqnarray}
where $t_i$ is the time spent in $A$ in one sojourn,
so it is distributed with $f_{\tau_E}(t_i)$ (\ref{ftau}).
Denoting $p_2$ the probability to hop to $B_2$ when starting at $B_1$
\begin{eqnarray}
p_2= \frac{w_{ B_{1} \to B_2} }{w_{ B_{1} \to B_2}
+ w_{ B_{1} \to A}} =
 \frac{ 1 }{1+ e^{+ \beta \alpha E}} \sim e^{- \beta \alpha E}
\end{eqnarray}
we may express the probability distribution $W(k)$ of $k=1,2,...$ as
\begin{eqnarray}
W(k)= (1-p_2)^{k-1} p_2
\end{eqnarray}
The Laplace transform of $t^*$ (\ref{tetoile}) thus reads
\begin{eqnarray}
\int_0^{+\infty} dt^* P_*(t^*) e^{- s t^*} 
= \sum_{k=1}^{+\infty} W(k) \left( \frac{1}{1+s \tau_E} \right)^k
= \frac{1}{1+s \frac{\tau_E}{p_2} }
\end{eqnarray}
so $t^*$ is exponentially distributed 
\begin{eqnarray}
 P_*(t^*)  
= \frac{1}{\tau^*_E} e^{- \frac{t^*}{\tau^*_E} }
\end{eqnarray}
with the mean time
\begin{eqnarray}
\tau^*_E = \frac{\tau_E}{p_2} = e^{ \beta  E}
\left( 1+e^{- \beta \alpha E} \right) \opsimeq e^{ \beta  E}
\label{tauetoile}
\end{eqnarray}
which is independent of the parameter $\alpha$, and exactly
coincides with the trapping time of the usual trap model ($\alpha=0$)
distributed with (\ref{qtau}).
In conclusion, as $\alpha$ grows from $0$ to $1$, the mean sojourn-time
$\tau_E$ (\ref{tauE}) in the trap A decays, but this is exactly
compensated by the large probability to return immediately back to $A$
when on nearest neighbor sites $B_{\pm 1}$.

\subsection{ Escape time
from a renormalized trap}

Since we have found that the trapping time $\tau^*_E$ (\ref{tauetoile}) of
the region $B_{-1} A B_{+1}$
containing a renormalized trap $A$ with its two neighbors
was exactly the same as the trapping time $\tau$ of the usual
trap model ($\alpha=0$) studied in the preceding Sections, 
we obtain that the escape properties from a renormalized trap
to another renormalized trap are exactly the same 
as for $\alpha=0$.
So the distribution of the escape time $T$ is given by 
(\ref{scalingqt}), where the RG scale $R(t)$ 
for the trapping times $\tau_E^*$ is related to the energy
RG scale $\Gamma(t)$ via (\ref{tauetoile})
\begin{eqnarray}
R(t) = e^{ \beta \Gamma(t)}
\label{defofgammat}
\end{eqnarray}

\subsection{ Conclusions in $d=1$ }

All results based on the escape times $T$
in the renormalized landscape do not depend on $\alpha$ :
in particular, the diffusion front, the localization parameters,
the disorder averages of thermal cumulants, and the correlation $C(t',t)$
do not depend on $\alpha$.
Among the observables we have considered, the only change
will be for the probability $\Pi(t_w+t,t_w)$ of no jump
between $t_w$ and $t_w+t$, since it directly involves
the distribution of trapping time $\tau_E$
(\ref{defpittprime}).
At lowest order in $\mu$ we have that the probability distribution
$\psi_t(\tau)$ to be in a trap of trapping time $\tau$ at time $t$
is now given by (\ref{qalpha})
\begin{eqnarray}
 \psi_{t}(\tau) = q^{(\alpha)}_{\Gamma} (\tau) = 
\theta( \tau> \tau_{\Gamma(t)} )
\frac{ \mu(\alpha) }{\tau} 
\left( \frac{\tau_{\Gamma(t)} } {\tau} \right)^{\mu(\alpha)}
\end{eqnarray}
where the cut-off 
is related to the RG scale $R(t)$ (\ref{roft}) via
(\ref{taugamma},\ref{tauetoile}))
\begin{eqnarray}
\tau_{\Gamma(t)} = e^{ \beta (1- \alpha ) \Gamma (t) }
= [R(t)]^{1-\alpha}
\end{eqnarray}

So we obtain the scaling form
\begin{eqnarray}
\Pi^{(\alpha)}_{\mu} (t+t_w,t) = {\tilde \Pi}_{\mu(\alpha)}^{(0)} \left( v= \frac{t}{[R(t_w)]^{1-\alpha}} \right)
\end{eqnarray}
in terms of the scaling function (\ref{piscal0})
with the modified exponent (\ref{mualpha}).

\subsection{ Effect of $\alpha$ in arbitrary dimension $d$ }

It is clear that the analysis presented above in $d=1$ 
can easily be generalized for the trap model
on a hypercubic lattice in arbitrary dimension $d$ :
a very deep trap $A$ of energy $E$ has a trapping time $\tau_E$ (\ref{tauE})
distributed with an algebraic law of exponent $\mu(\alpha)=\mu/(1-\alpha)$  
(\ref{mualpha}), whereas the cluster made out of $A$ and its $(2d)$
nearest neighbor sites has for trapping time  
$\tau^*_E \simeq e^{ \beta  E}$ (\ref{tauetoile})
distributed with an algebraic law of exponent $\mu(0)=\mu$.
This explains why the quenched model and the annealed model
remain different even in arbitrary dimension $d$ when $\alpha>0$,
as was found in the numerical simulations \cite{rinnmaassjp} :
in the quenched model, the exponent $\mu$ in the distribution of
$\tau^*_E$ comes from the very high probability to return
immediately to a deep trap when leaving it, and thus in the annealed
model where these returns are absent, the only relevant exponent
is $\mu(\alpha)=\mu/(1-\alpha)$.
As noted in \cite{rinnmaassjp},
this effect is particularly clear in the region $0<\mu<1<\frac{\mu}{1-\alpha}$  
 : the annealed model is above its glass transition,
whereas the quenched model is below its glass transition.

In conclusion, in the glassy phase $\mu<1$ in arbitrary dimension $d$, the diffusion front,
the localization parameters,
the disorder averages of thermal cumulants, and the correlation $C(t',t)$
do not depend on $\alpha$, and the only observables sensitive to $\alpha$
will be again those that measure the `local persistence', such
as the probability $\Pi(t_w+t,t_w)$ to remain exactly on the same site
between $t_w$ and $t_w+t$.
This conclusion is in agreement with
the qualitative argument given in Appendix B 3
of \cite{rinnmaassjp}.

\subsection{ Generalization to arbitrary 
jump rates satisfying detailed balance }

It has been recently argued in \cite{maassphilmag}
how the decay time for the function $\Pi(t_w+t,t_w)$
should depend on the form of the jump rates satisfying detailed balance 
\begin{eqnarray}
e^{\beta E_i} w(E_i  \ {\rm to}  \ E_j) = e^{\beta E_j} w(E_i \ {\rm to}
\  E_j)
\label{detailed}
\end{eqnarray}
when they are not of the special form (\ref{ratew}). 
It is thus interesting to consider this question from the RSRG perspective.

The arguments presented in details above for the 
special form (\ref{ratew}) of the rates 
may be straightforwardly generalized as follows.
The trapping time associated to a trap $A$ with energy
 $E> \Gamma$ existing in the renormalized landscape 
 (\ref{tauE}) is more generally given by
\begin{eqnarray}
\tau_E = \frac{1}{w(E \ {\rm to}  \ 0)}
\label{tauEgen}
\end{eqnarray}
whereas the trapping time of the vicinity
of this renormalized trap (\ref{tauetoile})
reads more generally
\begin{eqnarray}
\tau^*_E = \frac{\tau_E}{p_2} =
\frac{ w(0 \ {\rm to}  \ 0) + w(0 \ {\rm to}  \ E) }
{  w(E \ {\rm to}  \ 0) w(0 \ {\rm to}  \ 0) }
\label{tauetoilegen}
\end{eqnarray}
The detailed balance condition (\ref{detailed})
for the special case $E_i=E$ and $E_j=0$
thus gives
\begin{eqnarray}
\tau^*_E =
\frac{  e^{\beta E} }{ w(0 \ {\rm to}  \ 0) }+ \frac{ 1  }{  w(E \ {\rm to}  \ 0)  }
=\frac{  e^{\beta E} }{ w(0 \ {\rm to}  \ 0) }+ \tau_E
\label{tauetoiledet}
\end{eqnarray}
So the condition to have exactly the same properties on large scales
as in the usual trap model (\ref{ratew}) with $\alpha=0$, is simply
\begin{eqnarray}
\lim_{E \to \infty} (e^{-\beta E} \tau_E ) < +\infty
\end{eqnarray}
since this condition is sufficient to get $\tau^*_E \sim e^{\beta E}$.
In this case, the only observable sensitive to the precise form
of the rates will be again the probability $\Pi(t_w+t,t_w)$
 to remain exactly on the same site
between $t_w$ and $t_w+t$. Indeed, this probability will be a scaling function
of the ratio $(t/\tau_w(t_w))$ , where
$\tau_w(t_w)$ is the cut-off $\tau_{\Gamma(t_w)}$ of the renormalized landscape
at time $t_w$, i.e. at scale $R(t_w)$ (\ref{roft}).
In terms of the energy cut-off $\Gamma(t_w)$
(\ref{defofgammat})
\begin{eqnarray}
\Gamma(t_w) = \frac{ \ln R(t_w) }{ \beta } 
\sim \frac{ \ln t_w }{ \beta (1+\mu)  }
\end{eqnarray}
we thus obtain (\ref{tauEgen}) the general prediction
\begin{eqnarray}
\tau_w(t_w) \equiv \tau_{\Gamma(t_w)}
 = \frac{1}{w(\Gamma(t_w) \ {\rm to}  \ 0)} 
= \frac{1}{ w \left(\frac{ \ln t_w }{ \beta (1+\mu)  } \ {\rm to}  \ 0 \right)}
\label{taugtw}
\end{eqnarray}
Vice versa, if one is interested into
constructing a model with a given decay time decay time \cite{maassphilmag}
$\tau_w(t_w)$ for the function $\Pi(t_w+t,t_w)$, it is sufficient
to choose a form for the rates that satisfies
the following asymptotic behavior
\begin{eqnarray}
w(E \ {\rm to}  \ 0) 
\opsim_{E \to \infty} \frac{1}{\tau_w ( e^{\beta E (1+\mu) } )}
\end{eqnarray}
In particular, to obtain the ultra-slow logarithmic
aging dynamics $\tau_w(t_w)=(\ln t_w)^{\chi}$, the jump rates have 
to behave as the power law $w(E \ {\rm to}  \ 0) \sim E^{- \chi}$
for $E \to \infty$,
in agreement with \cite{maassphilmag}.

\section{ Scaling analysis for the trap model in dimension $d=2$ }

\label{dim2}

Since the random walk on a square lattice in dimension $d=2$ is recurrent,
we expect that here also, the scale of ``escape times"
in the renormalized landscape will be different from
the scale of trapping times. On the contrary, for $d>2$, the random
walk is not recurrent, and there will be only one time scale
given by the trappings times.

\subsection{ Renormalized landscape }

As before, we define the renormalized landscape at scale $R$
 as follows : all traps with trapping time $\tau_i<R$
are decimated and replaced by a ``flat landscape'', whereas 
all traps with waiting time $\tau_i>R$ remain unchanged. 

We now consider the probability $P_R(l)$ that 
the renormalized trap closest to the origin is at a distance $l$
from the origin.
At large scale, this means that the disc of surface $S=\pi l^2$
is empty of renormalized traps, whereas the disc of radius $l+dl$ is not,
yielding 
\begin{eqnarray}
  P_R (l) 
\opsimeq_{R \to \infty} \frac{2 \pi l}{R^{\mu} } e^{- 
\displaystyle \frac{ \pi l^2}{R^{\mu}} }
\label{lofrd2}
\end{eqnarray}
that generalizes (\ref{length},\ref{plambda}).
Given a renormalized trap $\tau_0$, $P_R(l)$ also describes
 the distribution of the distance $l$
of the nearest neighbor renormalized trap.

\subsection{ Escape time for a renormalized trap in $d=2$ }

We thus define the escape from a renormalized trap $\tau_0$
as the event where a particle starting at $\tau_0$ 
reaches for the first time the circle of distance $l$ of the nearest
renormalized trap : of course, contrary to the one-dimensional case,
this does not mean that the particle has been really absorbed, but
we expect that the particle has then a finite probability
to be absorbed by a renormalized trap different from $\tau_0$.

The probability to escape to the distance $l$ without returning to $\tau_0$
when starting nearby behaves as $c/(\ln l)$, where $c$ is a constant (see Appendix  
\ref{appexcursionsd2} for more details), and thus the number $n$
of sojourns at $\tau_0$ before the particle succeeds to escape 
scales as
\begin{eqnarray}
n \sim \ln l \sim \ln R^{\frac{\mu}{2} }
\end{eqnarray}
So the time spent inside the trap $\tau_0$ before the escape scales as
 \begin{eqnarray}
t_{in} \sim n \tau_0 \sim \tau_0 \ln R^{\frac{\mu}{2} } \geq 
R \ln R^{\frac{\mu}{2} }
\label{tind2}
\end{eqnarray}

The mean time $<t>_l$ of one unsuccessful excursion scales as
(see Appendix  
\ref{appexcursionsd2} for more details)
\begin{eqnarray}
<t>_l
\opsimeq_{l \to \infty} c' \frac{ l^2}{  \ln^2 l} 
\end{eqnarray}
so that the total time $t_{out}$ during the unsuccessful excursions scales 
before the escape scales as
\begin{eqnarray}
t_{out}
\sim n <t>_l \sim \frac{ l^2}{  \ln l} \sim \frac{ R^{\mu} }{ \ln R^{\frac{\mu}{2} } }
\end{eqnarray}
It is thus negligible with respect to $t_{in}$ (\ref{tind2})
at large scale.

Finally, the time $t_{diff}$ of the successful excursion
over the length $l$ scales as
 \begin{eqnarray}
t_{diff}
\sim  l^2 \sim  R^{\mu} 
\end{eqnarray}
and is thus also negligible with respect to $t_{in}$ (\ref{tind2})
at large scale.

In conclusion, the total time needed to escape from a trap $\tau_0$
of the renormalized landscape at scale $R$ 
 \begin{eqnarray}
t_{esc} = t_{in} +t_{out} +t_{diff} 
\end{eqnarray}
is dominated by $t_{in}$ alone at large scale, and thus 
we may associate to the trap $\tau_0$ an escape time 
 \begin{eqnarray}
T_0 = a \tau_0 \ln R^{\frac{\mu}{2} } 
\label{escaped2}
\end{eqnarray}
where $a$ is a random variable of order 1 that 
characterizes the geometry of the neighboring renormalized traps
around $\tau_0$. 

\subsection{Choice of the renormalization scale $R$ as a function of time}

The distribution of the escape time $T$ (\ref{escaped2})
in the renormalized landscape
at scale $R$ will thus present the scaling form
\begin{eqnarray}
Q_R(T) = 
 \frac{1}{R \ln R^{\frac{\mu}{2} }} {\hat Q}_{\mu} \left(
{\tilde T} = \frac{T}{R \ln R^{\frac{\mu}{2} } } \right)
\end{eqnarray}
So the renormalization scale $R$ has to be chosen as a function of time
with the scaling
\begin{eqnarray}
t \sim R(t) \ln R^{\frac{\mu}{2} }(t) 
\end{eqnarray}
i.e. by inversion at leading order
\begin{eqnarray}
R(t) \sim \frac{ t}{ \ln t^{\frac{\mu}{2} } }
\label{rdetd2}
\end{eqnarray}

So the recurrence properties of the random walk
in $d=2$ induces again a different scaling for
trapping times and escape times in the renormalized
landscape. As a consequence,
the correlation $\Pi(t+t_w,t_w)$ will be a scaling function of $t/R(t_w)$
with (\ref{rdetd2}) whereas the correlation $C(t+t_w,t_w)$ 
will be a scaling function of $t/t_w$, 
in agreement with the results obtained in 
\cite{cerny,reviewbenarous}.

\subsection{ Number of important traps}

However, in contrast with the $d=1$ case, there are two important
length scales in $d=2$ as we now explain. 

On one hand, given the choice (\ref{rdetd2})
 for the renormalization scale $R(t)$
as a function of time,
the length scale for the distance
between two renormalized traps is then given by (\ref{lofrd2})
\begin{eqnarray}
\xi(t) \sim [R(t)]^{\frac{\mu}{2} } \sim 
\left[ \frac{ t}{ \ln t^{\frac{\mu}{2} } } \right]^{\frac{\mu}{2} }
\label{xid2}
\end{eqnarray}

On the other hand, let us now recall the
scaling analysis using L\'evy sums \cite{alexander,argumentlevy,rinnmaassjp}
in $d=2$ : after $N$ steps, the number of visited sites 
behaves as $S=N/(\ln N)$ and each site is typically visited $\ln N$ times,
which yields the correspondence 
\begin{eqnarray}
t \sim \sum_{i=1}^{S} \tau_i (\ln N) \sim
(\ln N) \left( \frac{N}{\ln N} \right) ^{\frac{1}{\mu}}
\label{tofN} 
\end{eqnarray}
Since the typical distance reached after $N$ steps 
scales as $r \sim \sqrt N $, this yields in terms of the time $t$
 after the inversion of (\ref{tofN}) at leading order
\begin{eqnarray}
r(t) \sim \frac{ t^{\frac{\mu}{2} } }{ ( \ln t^{\mu} )^{\frac{\mu-1}{2}} }
\label{roftd2} 
\end{eqnarray}
So here, in contrast with the $d=1$ case, the typical distance $r(t)$
reached after time $t$ does not coincide with 
the typical distance $\xi(t)$ (\ref{xid2}) between 
two renormalized traps. As a consequence, the number
$n(t)$ of important traps contained in the disk
of radius $r(t)$ will not remain finite but
will grow logarithmically in time 
\begin{eqnarray}
n(t) \sim \frac{ r^2(t)} {\xi^2(t)} \sim  \ln t^{\mu}
\label{numbertraps}
\end{eqnarray}
This is in agreement with the absence of 
localization in $d=2$ as $t \to \infty$ \cite{reviewbenarous}.

However, in the regime where $t \to \infty$ and $\mu \to 0$
with $t^{\mu}$ fixed (\ref{numbertraps}), 
we expect that the diffusion front in a given sample may
be described as a sum of delta peaks that are out of equilibrium
with respect to each other, i.e. that their weights 
depend on the geometry of the renormalized traps
in the region around the origin but not on their energies.
It would be of course very interesting to build 
some RSRG effective dynamics for $d=2$ case in this regime, 
but this goes beyond the scope of this paper.

\section{ Conclusion}

\label{conclusion}

We have studied in details the
 properties of the one-dimensional trap model
via a disorder-dependent 
real-space renormalization procedure.
The RSRG approach
provides the correct exponents for the relevant length scale
and the two relevant time scales in the full domain $0 \leq \mu <1$,
and then allows to compute scaling functions
in a systematic series expansion in $\mu$.
Since we have already summarized our main results 
for disorder averages of observables in
the Introduction \ref{mainres}, we will not describe them again
but instead discuss the physical meaning of our 
construction.

We have seen that at lowest order
in $\mu$, in a given sample,
the particle can be only on two sites, which are the
two nearest renormalized traps $M_+$ and $M_-$ 
existing at RG scale $R(t)$, and
that the weights of these traps are simply given by the probabilities
to reach one before the other one. This means that the dynamics
remains out of equilibrium forever : the weights are not given by Boltzmann
factors, they don't even depend on the energies of these two traps,
but they only depend on the ratio of the distances to the origin!
When including the first corrections in $\mu$ to this picture,
we have taken into account the biggest trap $S$ in the interval $]M_-,M_+[$,
and the next renormalized traps $M_{- -}$ and $M_{++}$, but the dynamics still remains out-of equilibrium since the weights are still
determined by the lengths between these traps and by escape rates
of the form $e^{-t/T}$. This out-of-equilibrium
character will thus persist in the whole localized phase $0 \leq \mu<1$
where these deep traps keep finite weights.  
This explains why the numerical studies \cite{rinnmaassjp,bertinjp}
have found that the ideas of ``partial equilibrium"
and ``effective temperature" were not able to capture 
the long-time properties of the trap model. 
This is similar to what happens in the directed trap model
or in the related biased Sinai diffusion, where the thermal packet
is also broken into sub-packets that remain out-of-equilibrium 
with each other even in the infinite time limit \cite{c_directed};
this is in contrast with the unbiased Sinai diffusion, where 
the thermal packet is asymptotically localized in one single infinite valley
\cite{golosov}, in which particles are at equilibrium 
which each other \cite{us_golosov}.

From the point of view of numerical simulations
on random walks in random media, 
this shows that to study the localization and the convergence or not to a quasi-equilibrium
regime at long times, it is interesting to study the dynamics
in a single disordered sample, and not only disorder-averaged quantities :
for instance, the pictures of a thermal packet in a given sample
obtained in \cite{chave} for the Sinai diffusion,
and in \cite{comptejpb} for the directed trap model,
seems to give the clearer insight into the question of localization
in one valley or in typically a few traps with finite weights. 
 And whenever the structure of the thermal packet
consists in a few sub-packets, whose positions and weights are
sample-dependent, it is a very strong indication that some 
appropriate real-space RG description can be constructed to study
the dynamics.
 
In a forthcoming paper \cite{c_trapreponse}, we will show
 how the present RSRG approach
for the unbiased trap model can be generalized to obtain
explicit results in the limit $\mu \to 0$
for the linear and non-linear 
response to an external bias, when it is applied from the very beginning
at $t=0$ or after a waiting time $t_w$.
Recently, this problem has been studied via scaling arguments
and numerical simulations in \cite{bertinjpreponse},
and was shown to satisfy a non-linear Fluctuation Theorem \cite{c_nonlinearft}.

\begin{acknowledgments}

It is a pleasure to thank E. Bertin, J.P. Bouchaud 
and J.M. Luck for useful discussions, as well as J. Cerny
for sending me his PhD Thesis \cite{cerny}.

\end{acknowledgments}

\appendix

\section{ Statistical properties of excursions in $d=1$ }

\label{appexcursions}

As explained in the text, to study the excursions in the 
renormalized landscape
(\ref{tout},\ref{tdiff}), we have to study the following standard problem :
what is the probability distribution
$P_x(t)$ of the time $t$ of the first-passage at $x=0$
without having touched the other boundary $x=l$ before,
for a particle in a pure trap model?

For $x=1,...,l-1$ this probability distribution satisfies the equation
\begin{eqnarray}
\partial_t P_x(t) =  \left[ P_{x+1}(t) + P_{x-1}(t) -2 P_x(t) \right]
\end{eqnarray}
with the boundary conditions $P_0(t)= \delta(t)$ and $P_l(t)=0$.
So the Laplace transform with respect to $t$
\begin{eqnarray}
{\hat P}_x(s) \equiv \int_0^{+\infty} dt e^{-s  t }  P_x(t)
\end{eqnarray}
satisfies
\begin{eqnarray}
{\hat P}_{x+1}(s) + {\hat  P}_{x-1}(s) 
- (2+s) {\hat  P}_x(s) = 0
\end{eqnarray}
for $x=1,...,l-1$
with the boundary conditions 
${ \hat P}_0(s)= 1$ and ${\hat P}_l(s)=0$
The solution reads
\begin{eqnarray}
 {\hat  P}_x(s) && =  \frac{ \rho^x(s) - \rho^{2l-x}(s) }{1- \rho^{2l}(s)} \\
\rho(s) && = \frac{2+s - \sqrt{s^2+4s} }{2}
\end{eqnarray}

In particular, the series expansion in $s$ yields the first moments
\begin{eqnarray}
\theta_k(x) \equiv \int_0^{+\infty} dt t^k P_x(t)
\label{moments}
 \end{eqnarray}
For $n=0$, the probability to reach $0$ before $l$ when starting at $x$
reads as expected
\begin{eqnarray}
\theta_0(x) = \frac{l-x}{l}
\end{eqnarray}
For $n=1,2$
\begin{eqnarray}
\theta_1(x) && = \frac{x (l-x) (2l-x)}{6 l}  \\
\theta_2(x) && = \frac{x (l-x) (2l-x) (4 l^2+6 x l - 3 x^2 +5) }{360 l}
\end{eqnarray}

\subsubsection{ Unsuccessful excursions }

For the unsuccessful excursions present in $t_{out}$ (\ref{tout}),
we need to consider the special case $x=1$,
where the first moments read
\begin{eqnarray}
\theta_0(1) && = 1- \frac{1}{l} \\
\theta_1(1) && = \frac{ (l-1) (2l-1)}{6 l} \opsimeq_{ l \to \infty} \frac{l}{3} \\
\theta_2(1) && = \frac{ (l-1) (2l-1) (4 l^2+6  l - 3  +5) }{360 l}
\opsimeq_{ l \to \infty} \frac{l^3}{45} 
\end{eqnarray}
The leading terms can be understand as follows :
the time $t$ is of order $l^2$ with a probability $1/l$,
which is the probability to arrive very near the
forbidden extremity $x=l$.

To get the leading behavior in $l$ of all moments,
we thus need to consider the rescaled Laplace transform
\begin{eqnarray}
 {\hat  P}_1 \left( \frac{p}{l^2} \right) \opsimeq_{l \to \infty} 
= 1- \frac{1}{l} (\sqrt p \coth \sqrt p -1 )  +O\left(\frac{1}{l^2} \right) \end{eqnarray}
which corresponds to the following
 leading behavior at large $l$ for the moments
\begin{eqnarray}
\theta_k(1) 
\opsimeq_{l \to \infty} l^{2k-1} \frac{2 \Gamma(k+1) \zeta(2k) }{\pi^{2 k} } 
\end{eqnarray}

\subsubsection{ Successful excursions }

For the successful excursion present in $t_{diff}$ (\ref{tdiff}),
we need to consider the special case $x=l-1$.
Since in this case, the normalization is
\begin{eqnarray}
\theta_0(l-1) = \frac{1}{l}
\end{eqnarray}
we consider the `normalized' first moments 
\begin{eqnarray}
 \frac{\theta_1(l-1)}{\theta_0(l-1)}  
&& \opsimeq_{l
\to \infty} \frac{l^2}{6}  \\
\frac{\theta_2(l-1)}{\theta_0(l-1)} 
&& 
 \opsimeq_{l
\to \infty} \frac{ 7 l^4  }{360 }
\end{eqnarray}
To get the leading behavior in $l$ of all moments,
we thus need to consider the rescaled Laplace transform
\begin{eqnarray}
 {\hat  P}_{l-1} \left( \frac{p}{l^2} \right) \opsimeq_{l \to \infty} 
=  \frac{1}{l} \frac{ \sqrt p}{ \sinh \sqrt p }  +O\left(\frac{1}{l^2} \right) \end{eqnarray}
So the time $t_{diff}$ for the successful diffusion
scales as $l^2$ as expected.

\section{ Statistical properties of excursions in $d=2$ }

\label{appexcursionsd2}

To study the scaling properties of
excursions in the 
renormalized landscape in dimension $d=2$, we have to study the following problem :
what is the probability distribution
$P_r(t)$ of the time $t$ of the first-passage 
on the circle $r=1$
without having touched the other circle $r=l$ before,
for a particle starting at radius $r$ and diffusing freely?

For $1<r<R$, this probability distribution satisfies the diffusion equation
in radial coordinates
\begin{eqnarray}
\partial_t P_r(t) =  \Delta P_r(t) = \left( \frac{d^2}{dr^2}
+ \frac{ 1}{r} \ \frac{ d}{d r} \right) P_r(t) 
\end{eqnarray}
with the boundary conditions $P_1(t)= \delta(t)$ and $P_l(t)=0$.

So the Laplace transform with respect to $t$
\begin{eqnarray}
{\hat P}_r(s) \equiv \int_0^{+\infty} dt e^{-s  t }  P_r(t)
\end{eqnarray}
satisfies
\begin{eqnarray}
 \left( \frac{d^2}{dr^2}
+ \frac{ 1}{r} \ \frac{ d}{d r} -s \right) P_r(s) =0
\end{eqnarray}
for $1<r<R$
with the boundary conditions 
${ \hat P}_0(s)= 1$ and ${\hat P}_l(s)=0$.

In particular, the moments appearing in the series expansion 
of $P_r(s)$ in $s$ 
\begin{eqnarray}
\theta_k(r) \equiv \int_0^{+\infty} dt t^k P_r(t)
\label{momentsd2}
 \end{eqnarray}
may be computed by the recurrence
\begin{eqnarray}
 \left( \frac{d^2}{dr^2}
+ \frac{ 1}{r} \ \frac{ d}{d r}  \right) \theta_k(r)
= - \theta_{k-1}(r)
\end{eqnarray}

For $k=0$, the probability $\theta_0(r)$ to reach first $r=1$
before $r=l$ when starting at $r$ is the solution of
\begin{eqnarray}
 \left( \frac{d^2}{dr^2}
+ \frac{ 1}{r} \ \frac{ d}{d r}  \right) \theta_0(r)
= 0
\end{eqnarray}
with the boundary conditions 
$\theta_0(1)= 1$ and $\theta_0(l)=0$, and thus reads
\begin{eqnarray}
 \theta_0(r)
= 1 -\frac{ \ln r}{\ln l}
\end{eqnarray}

For $k=1$, $\theta_1(r)$ is the solution of
\begin{eqnarray}
 \left( \frac{d^2}{dr^2}
+ \frac{ 1}{r} \ \frac{ d}{d r}  \right) \theta_1(r)
= - \left( 1 -\frac{ \ln r}{\ln l} \right)
\end{eqnarray}
with the boundary conditions 
$\theta_1(1)= 0$ and $\theta_1(l)=0$, and thus reads
\begin{eqnarray}
 \theta_1(r)
= \frac{ l^2-1}{ 4 \ln^2 l} \ln r + \frac{ (r^2-1) ( \ln \frac{r}{l} -1)}
{ 4 \ln l} 
\end{eqnarray}

In particular, to study the unsuccessful excursions, we may consider $r=2$
for starting point. The probability to escape is then
\begin{eqnarray}
 \theta_0(2)
 = 1 -\frac{ \ln 2}{\ln l} 
\end{eqnarray}
and the mean time of an unsuccessful excursion has for leading behavior in
$l$
\begin{eqnarray}
 \theta_1(2)
\opsimeq_{l \to \infty}  \frac{ l^2}{ 4 \ln^2 l} \ln 2 
\end{eqnarray}

\section{ Effective model at first order in $\mu$ }

\label{correcmu1}

In this Appendix, we compute the corrections at first order in $\mu$
by taking into account the two effects described in the text 
in Section \ref{dispersionmu}

\subsection{Statistical properties for particles ``in advance" at order $\mu$}

There is a probability 
\begin{eqnarray}
\pi_a=(1-e^{-\frac{t}{T_{M}} })
\end{eqnarray}
that a particle has already escape at time $t$
from a renormalized trap of escape time $T_M$.

We consider a particle starting at $O$
that was in the trap $M_+$
in the effective dynamics of previous Section.
With probability $\pi_a$,
the particle has already escaped from the trap $M_+$ at time $t$,
and then it has been absorbed by $M_-$ or by $M_{++}$,
the two nearest renormalized traps,
with probabilities given by the ratios of the distances.

We note as before $X_{\pm}$ the rescaled distances between the origin and $M_{\pm}$, and we note $\lambda$ the rescaled distance between $M_+$ and $M_{++}$. 
The joint probability distribution of $(X_-,X_+,\lambda,\tau_M)$ 
is completely factorized
 \begin{eqnarray}
{\cal D}_{M_-M_+M_{++}}( X_-,X_+,\lambda,\tau_M) =
 \theta(X_+) \theta(X_-) \theta(\lambda) \theta(\tau_M>R(t))
\mu \frac{d\tau_M}{\tau_M} \left( \frac{R(t)}{\tau_M} \right)^{\mu}
 e^{-X_+ -X_- -\lambda} 
\label{measureMMM++one}
\end{eqnarray}

Since the escape time $T_M$ is proportional to the trapping time
 \begin{eqnarray}
T_M= \frac{1}{a_M} R^{\mu} \tau_M
\end{eqnarray}
with a prefactor $a_M$ that depends on the positions via
the prefactor
 \begin{eqnarray}
a_M=  \frac{1}{2} \left(
\frac{1}{ (X_+ +X_-) }+ \frac{1}{  \lambda } \right) 
\end{eqnarray}
it is more convenient to use the variable
 \begin{eqnarray}
\lambda_0 && =X_- +X_+
\end{eqnarray}
instead of $X_+$.
So the measure (\ref{measureMMM++one}) becomes
 \begin{eqnarray}
{\cal D}_{M_-M_+M_{++}}( X_-,\lambda_0,\lambda,\tau_M) =
 \theta(0<X_-<\lambda_0) \theta(\lambda) \theta(\tau_M>R(t))
\mu \frac{d\tau_M}{\tau_M} \left( \frac{R(t)}{\tau_M} \right)^{\mu}
 e^{- \lambda_0  -\lambda} 
\label{measureMMM++}
\end{eqnarray}

To compute the average of an observable $A(X_-,\lambda_0,\lambda,\tau_M)$
 with respect to this measure, 
it is more convenient to first integrate over $X_- \in (0,\lambda_0)$
and then to integrate over the remaining variables with the following notations
 \begin{eqnarray}
\overline{A(X_-,\lambda_0,\lambda,\tau_M)} = << \int_0^{\lambda_0} dX_- A(X_-,\lambda_0,\lambda,\tau_M) >>_a
\label{firststep}
\end{eqnarray}
where the notation $<<..>>_a$ denotes
 \begin{eqnarray}
<< f(\lambda_0,\lambda,\tau_M) >>_a
= \int_{R(t)}^{+\infty} d\tau_M  \mu \frac{d\tau_M}{\tau_M} \left( \frac{R(t)}{\tau_M} \right)^{\mu}
\int_0^{+\infty} d \lambda_0
\int_0^{+\infty} d \lambda
 e^{- \lambda_0  -\lambda} f(\lambda_0,\lambda,\tau_M)
\label{meanwitha}
\end{eqnarray}

For the simplest observables, we will need
integrals of the following form, which can be computed 
in terms of Bessel Functions via (\ref{inteleadingtoK})
 \begin{eqnarray}
&& W_{p,q} (\mu) \equiv 
<< (1- e^{- \frac{t}{2 R^{\mu} \tau_M}
\left(  \frac{1}{\lambda} + \frac{1}{\lambda_0} \right) } ) \lambda_0^p \lambda^q  >>_a \nonumber \\
&& = 2 \mu \int_0^{z_0(\mu)} \frac{dz}{z} 
\left( \frac{z}{z_0(\mu)}\right)^{2 \mu}
\left[ 
\Gamma(1+p) \Gamma(1+q) - \frac{z^{2+p+q}}{  2^{p+q}}
 K_{1+p}(z)
K_{1+q}(z) \right]
\label{wpq}
\end{eqnarray}
where the parameter $z_0$ depends only of $\mu$ via
${\tilde T}_0$ introduced in the explicit choice of the renormalization scale $R(t)$ (\ref{roft})
 \begin{eqnarray}
z_0(\mu)
\equiv \sqrt{ \frac{2t}{R^{1+\mu}(t) } } 
= \sqrt{ 2 {\tilde T}_0 (\mu) }
\label{defz0mu}
\end{eqnarray}

In particular, in the limit of vanishing $\mu$, 
$z_0(\mu)$ has for limit (\ref{tildet0mu0})
 \begin{eqnarray}
z_0(0)
=   2 e^{ - \frac{(1+\gamma_E)}{2} } =0.90895
\label{z0mu0}
\end{eqnarray}
so that the integrals read at lowest order in $\mu$
 \begin{eqnarray}
 W_{p,q} (\mu)  \opsimeq_{\mu \to 0}
 2 \mu \int_0^{z_0(0)} \frac{dz}{z} 
\left[ 
\Gamma(1+p) \Gamma(1+q) - \frac{z^{2+p+q}}{  2^{p+q}}
 K_{1+p}(z)
K_{1+q}(z) \right]
 + O(\mu^2)
\label{wpqmu0}
\end{eqnarray}

As first example, let us consider the probability to have already
escaped from a renormalized trap at time $t$
 \begin{eqnarray}
\pi_a  =  1- e^{- \frac{t}{T_M } } 
= 1- e^{- \frac{t}{2 R^{\mu} \tau_M } \left(\frac{1}{ \lambda_0 }+ \frac{1}{  \lambda } \right) }
\end{eqnarray}
Its average over the configurations can be computed via
 \begin{eqnarray}
\overline{\pi_a}  = << \lambda_0 
\left[ 1- e^{- \frac{t}{2 R^{\mu} \tau_M } \left( \frac{1}{ \lambda_0 }+ \frac{1}{  \lambda } \right) } \right] >>_a
\end{eqnarray}
which corresponds to the form (\ref{wpq}) 
for the special case $(p=1,q=0)$
and is thus of order $\mu$ (\ref{wpqmu0})
 \begin{eqnarray}
\overline{\pi_a}  = W_{1,0} (\mu) \opsimeq_{\mu \to 0}
 2 \mu \int_0^{2 e^{ - \frac{(1+\gamma_E)}{2} }} \frac{dz}{z} 
\left[ 1 - \frac{z^{3}}{  2}
 K_{2}(z)
K_{1}(z) \right]
 + O(\mu^2) = \mu \ 0.678238  + O(\mu^2)
\end{eqnarray}

Let us now write the diffusion front
in a given sample :
it is a linear combination of the two
possibilities (\ref{rulespplusmoins},\ref{ruledecim})
\begin{eqnarray}
&&  {\cal P}^{(0)+(1)}_{M_- M_+ M_{++}} (n) = 
e^{-\frac{t}{T_{M}} } \left[ 
p_{[M_-M_+]}(M_- \vert 0) \delta_{n,n_{M_-}}
+ p_{[M_-M_+]}(M_+ \vert 0) \delta_{n,n_{M_+}} \right] \nonumber \\
&& + (1-e^{-\frac{t}{T_{M}} }) 
\left[ p_{[M_-M_{++}]}(M_- \vert 0) \delta_{n,n_{M_-}}
+ p_{[M_-M_{++}]}(M_{++} \vert 0) \delta_{n,n_{M_++}} \right] 
\end{eqnarray}
and thus in rescaled distances, the 
correction with respect to the zero-th order
(\ref{pMM}) reads
\begin{eqnarray}
&& {\cal P}^{(1)}_{M_- M_+ M_{++}} \left( X  \right)
 \equiv  {\cal P}^{(0)+(+1)}_{M_-,M_+,M_{++}} \left( X  \right)
-  {\cal P}^{(0)}(X)  \nonumber \\
&&  = (1-e^{-\frac{t}{T_{M}} }) 
 \left[ -  \frac{X_-}{\lambda_0}  \delta(X-(\lambda_0-X_-)) 
+   \frac{\lambda X_-}{\lambda_0(\lambda+\lambda_0)} \delta(X+X_-) 
+  \frac{X_-}{\lambda+\lambda_0} \delta(X-(\lambda+\lambda_0-X_-)) 
 \right]
 \label{pMMM++}
\end{eqnarray}
where the parameters are distributed with the measure
(\ref{measureMMM++}).

Similarly, if the particle has escaped from the $M_-$, it has been absorbed
by $M_+$ or by $M_{- -}$ and the properties are the same as above 
by the symmetry $X \to -X$.

\subsection{Statistical properties of particles ``in delay" }

We are now interested into the trap $S$ with the
biggest trapping time $\tau_S$ in the interval $]M_-,M_+[$.
Its position is uniformly distributed in this interval,
so the joint distribution of $(X_-,X_S,X_+,\tau_S)$
reads
 \begin{eqnarray}
{\cal D}_{M_-SM_+}( X_-,X_S,X_+,\tau_S) =
 \theta(X_+) \theta(X_-) \theta(-X_-<X_S<X_+) \theta(\tau_S<R(t))
\frac{\mu}{\tau_S} \left( \frac{R(t)}{\tau_S} \right)^{\mu}
 e^{-(X_+ +X_-) \left( \frac{R(t)}{\tau_S} \right)^{\mu} } 
\label{measureM-SMM+one}
\end{eqnarray}
The measure for the positions alone reads
 \begin{eqnarray}
\int d\tau_s {\cal D}_{M_-SM_+}( X_-,X_S,X_+,\tau_S) =
 \theta(X_+) \theta(X_-) \theta(-X_-<X_S<X_+) 
\frac{1}{X_++X_-} e^{-(X_+ +X_-) } 
\end{eqnarray}
whereas the measure for the trapping time alone reads
 \begin{eqnarray}
\int dX_+ \int dX_- \int dX_S {\cal D}_{M_-SM_+}( X_-,X_S,X_+,\tau_S) =
\theta(\tau_S<R(t)) 2 \mu \frac{\tau_s^{2 \mu-1} }{R^{2 \mu}(t) } 
\end{eqnarray}

Since the associated escape time 
 \begin{eqnarray}
T_S= \frac{1}{a_S} R^{\mu} \tau_S
\end{eqnarray}
 depends on the positions via
the prefactor
 \begin{eqnarray}
a_S=  \frac{1}{2} \left(
\frac{1}{ (X_+-X_S) }+ \frac{1}{  (X_-+X_S)} \right) 
\end{eqnarray}
it is more convenient to replace the rescaled distances
$(X_-,X_+)$ between $M_{\pm}$ and the origin by the rescaled distances
between $M_{\pm}$ and the trap $S$
 \begin{eqnarray}
\lambda_+ && =X_+-X_S \\
\lambda_- && =X_-+X_S
\end{eqnarray}
With these new variables, the measure (\ref{measureM-SMM+one})
becomes
 \begin{eqnarray}
{\cal D}_{M_-SM_+}(X_S,\tau_S;\lambda_+,\lambda_-) =
 \theta(\lambda_+) \theta(\lambda_-) \theta(-\lambda_+<X_S<\lambda_-) \theta(\tau_S<R(t))
\frac{\mu}{\tau_S} \left( \frac{R(t)}{\tau_S} \right)^{\mu}
 e^{-(\lambda_+ + \lambda_- ) \left( \frac{R(t)}{\tau_S} \right)^{\mu} } 
\label{measureM-SMM+}
\end{eqnarray}

To compute the average of an observable $A(X_S,\lambda_+,\lambda_-,\tau_S)$
 with respect to this measure, 
it is convenient to first integrate over $X_S \in [-\lambda_+,\lambda_-]$
 and then average over the remaining variables
 \begin{eqnarray}
\overline{A(X_S,\lambda_+,\lambda_-,\tau_S)} = 
<< \int_{-\lambda_+}^{\lambda_-} dX_S A(X_S,\lambda_+,\lambda_-,\tau_S) >>_d
\label{firststep2}
\end{eqnarray}
where the notation $<<..>>_d$ is defined by
 \begin{eqnarray}
<< f(\lambda_+,\lambda_-,\tau_S) >>_d
= \int_0^{R(t)} d \tau_S \frac{\mu  }{\tau_S} \left( \frac{R(t)}{\tau_S} \right)^{\mu}
\int_0^{+\infty} d \lambda_+
\int_0^{+\infty} d \lambda_-
 e^{-(\lambda_+ + \lambda_- ) \left( \frac{R(t)}{\tau_S} \right)^{\mu} }
f(\lambda_+,\lambda_-,\tau_S) 
\label{meanwithd}
\end{eqnarray}

For the simplest observables, we will need
integrals of the following form, which can be computed 
in terms of Bessel Functions via (\ref{inteleadingtoK})
 \begin{eqnarray}
&& \Omega_{p,q} (\mu) \equiv << e^{- \frac{t}{T_S} } \lambda_+^p \lambda_-^q  >>_d 
 = \frac{2 \mu}{1+\mu} \frac{z_0^{\frac{2 \mu}{1+\mu} (1+p+q)}(\mu) }{2^{p+q}}
\int_{z_0(\mu)}^{+\infty} dz z^{\frac{1- \mu}{1+\mu} (1+p+q)} K_{1+p}(z)
K_{1+q}(z)
\label{omegapq}
\end{eqnarray}
where the parameter $z_0$ has been introduced in (\ref{defz0mu}).
At lowest order in $\mu$ we thus obtain
 \begin{eqnarray}
&& \Omega_{p,q} (\mu) \opsimeq_{\mu \to 0}
 \frac{  2 \mu }{2^{p+q}}
\int_{z_0(0)}^{+\infty} dz z^{ (1+p+q)} K_{1+p}(z)
K_{1+q}(z) + O(\mu^2)
\label{omegapqmu0}
\end{eqnarray}

Let us first consider the probability 
\begin{eqnarray}
\pi_d=e^{-\frac{t}{T_{S}} } =e^{-\frac{t}{ 2 R^{\mu} \tau_{S}}
\left(
\frac{1}{ \lambda_+ }+ \frac{1}{  \lambda_-} \right) } 
\end{eqnarray}
that a particle that has been trapped by $S$
is still in the trap $S$ at time $t$.
Its average 
with respect to the measure (\ref{measureM-SMM+}) can be computed as
\begin{eqnarray}
 \overline{\pi_d} = << (\lambda_+ + \lambda_-)
e^{-\frac{t}{ 2 R^{\mu} \tau_{S}}
\left(
\frac{1}{ \lambda_+ }+ \frac{1}{  \lambda_-} \right) } >>_d
 \end{eqnarray}

It is of the form (\ref{omegapq})
 and is thus of order $\mu$ (\ref{omegapqmu0}) 
\begin{eqnarray}
 \overline{\pi_d} = 2 \Omega_{1,0}(\mu) \opsimeq_{\mu \to 0} && 2 \mu  
\int_{2 e^{ - \frac{(1+\gamma_E)}{2} }}^{+\infty} dz z^2 K_{2}(z)
K_{1}(z) + O(\mu^2) 
  \opsimeq_{\mu \to 0} \mu 1.20205... +O(\mu^2)
\label{pid}
 \end{eqnarray}

Let us now write the diffusion front in a given sample at this order
\begin{eqnarray}
&&  {\cal P}^{(0)+(1)}_{M_- S M_+ } \left( X  \right) = 
e^{-\frac{t}{T_{S}} } \theta(X_S>0) \left[ 
p_{[M_-S]}(M_- \vert 0) \delta_{n,n_{M_-}}
+ p_{[M_-S]}(S \vert 0) \delta_{n,n_{S}} \right]  \nonumber \\
&& + e^{-\frac{t}{T_{S}} } \theta(X_S<0) \left[ 
p_{[SM_+]}(M_+ \vert 0) \delta_{n,n_{M_+}}
+ p_{[SM_+]}(S \vert 0) \delta_{n,n_{S}} \right]  \nonumber \\
&& + (1-e^{-\frac{t}{T_{S}} }) 
\left[ p_{[M_-M_{+}]}(M_- \vert 0) \delta_{n,n_{M_-}}
+ p_{[M_-M_+]}(M_{+} \vert 0) \delta_{n,n_{M_+}} \right] 
\end{eqnarray}
and thus in rescaled distances, the 
correction with respect to the zero-th order
(\ref{pMM}) reads
\begin{eqnarray}
&& {\cal P}^{(1)}_{M_- S M_+ } \left( X  \right)
 \equiv  {\cal P}^{(0)+(1)}_{M_- S M_+,} \left( X  \right)
-  {\cal P}^{(0)}(X)  \nonumber \\ 
&& = e^{-\frac{t}{T_{S}} } \theta(X_S>0) \left[ 
\frac{X_S}{\lambda_-} \delta(X+(\lambda_- -X_S))
+ \frac{(\lambda_- -X_S)}{\lambda_-} \delta(X-X_S) \right] \nonumber \\
&& + e^{-\frac{t}{T_{S}} } \theta(X_S<0) \left[ 
\frac{(-X_S)}{\lambda_+}  \delta(X-(\lambda_+ +X_S))
+ \frac{\lambda_+ + X_S}{\lambda_+} \delta(X-X_S) \right] \nonumber \\
&& -e^{-\frac{t}{T_{S}} } 
\left[ \frac{\lambda_++X_S}{\lambda_+ + \lambda_-} \delta(X+\lambda_- -X_S)
+ \frac{\lambda_- -X_S}{\lambda_+ + \lambda_-} \delta(X-(\lambda_++X_S)) \right]  \label{pM-SM+}
\end{eqnarray}

where the parameters are distributed with the measure (\ref{measureM-SMM+}).

\subsection{ Correction of order $\mu$ to the averaged diffusion front}

\subsubsection{Contribution of particles in advance}

To compute the contribution (\ref{pMMM++}) to the diffusion front of
the configurations $M_-M_+M_{++}$, we first integrate over 
$X_- \in [0,\lambda_0]$

\begin{eqnarray}
&& \int_0^{\lambda_0} dX_-
{\cal P}^{(1)}_{M_- M_+ M_{++}} \left( X  \right)
 = (1-e^{-\frac{t}{T_{M}} }) \nonumber \\
&& \times \left[ -  \frac{\lambda_0-\vert X \vert }{\lambda_0}  \theta(0<X<\lambda_0) 
+   \frac{\lambda \vert X \vert}{\lambda_0(\lambda+\lambda_0)} 
\theta(0<-X<\lambda_0) 
 +  \frac{\lambda + \lambda_0-\vert X \vert }{\lambda+\lambda_0} \theta(\lambda<X<\lambda+\lambda_0) 
 \right]
\end{eqnarray}

So taking into account the configurations $M_{--}M_-M_+$
via the symmetry $X \to -X$, we finally get that
averaging with the measure (\ref{measureMMM++})
yields the following contribution of particles in advance
to the scaling function at order $\mu$

\begin{eqnarray}
 {g}^{(1)}_{a} \left( X  \right)
&& = \mu \int_0^{+\infty} d \lambda_0 \int_0^{+\infty} d \lambda 
\int_0^{1} \frac{dw}{w}
 e^{-(\lambda_0 + \lambda )  } 
(1- e^{-\frac{ {\tilde T}_0(0) }{2} w \left( \frac{1}{\lambda_0} 
+ \frac{1}{\lambda} \right) } )  \nonumber \\
&& \left[-  2 \frac{\lambda_0-\vert X \vert }{\lambda_0}  \theta(0<X<\lambda_0) 
 +  \frac{\lambda + \lambda_0-\vert X \vert }{\lambda+\lambda_0} \theta(0<X<\lambda+\lambda_0)
\right]+)(\mu^2)
\label{g1a}
\end{eqnarray}

\subsubsection{Contribution of particles in delay}

To compute the average of the specific contribution (\ref{pM-SM+}) of the configurations
$S_-MS_+$  with respect to the measure (\ref{measureM-SMM+}),
we first integrate over $X_S$ 
\begin{eqnarray}
&& \int dX_S \theta(-\lambda_+<X_S<\lambda_-)
{\cal P}^{(1)}_{M_- S M_+ } \left( X  \right)
 \nonumber  \\ 
&& = e^{-\frac{t}{T_{S}} }  \left[ 
 \theta( \vert X \vert <\lambda_-)
\frac{ \lambda_-  - \vert X \vert }{\lambda_-} 
+ \theta( \vert X \vert <\lambda_+)
\frac{(\lambda_+ - \vert X \vert )}{\lambda_+}
- \theta(\vert X \vert <\lambda_+ + \lambda_-)
\frac{\lambda_- + \lambda_+ - \vert X \vert }{\lambda_+ + \lambda_-}
 \right] 
\end{eqnarray}
so that the correction to the averaged diffusion front reads
\begin{eqnarray}
&& g^{(1)}_{d} \left( X  \right)
 = \mu \int_0^{+\infty} d \lambda_+ \int_0^{+\infty} d \lambda_- 
\int_1^{+\infty} \frac{dw}{w}
 e^{-(\lambda_+ + \lambda_- ) } 
 e^{-\frac{ {\tilde T}_0(0) }{2} w \left( \frac{1}{\lambda_+} 
+ \frac{1}{\lambda_-} \right) }  \nonumber \\
&& \left[ 
 \theta( \vert X \vert <\lambda_-)
\frac{ \lambda_-  - \vert X \vert }{\lambda_-} 
+ \theta( \vert X \vert <\lambda_+)
\frac{(\lambda_+ - \vert X \vert )}{\lambda_+}
- \theta(\vert X \vert <\lambda_+ + \lambda_-)
\frac{\lambda_- + \lambda_+ - \vert X \vert }{\lambda_+ + \lambda_-}
 \right]+O(\mu^2) 
\label{g1d}
\end{eqnarray}

\subsubsection{Total correction at first order in $\mu$}

The two contributions (\ref{g1a},\ref{g1d})
yields the following total contribution at first order in $\mu$

\begin{eqnarray}
{g}^{(1)} \left( X  \right) && \equiv {g}^{(1)}_{a} \left( X  \right)
+ g^{(1)}_{d} \left( X  \right) 
 = \mu  \int_0^{+\infty} d \lambda_0 \int_0^{+\infty} d \lambda 
e^{-(\lambda_0 + \lambda )  } \left[ \gamma_E+ \ln \frac{ {\tilde T}_0(0) }{2} 
 \left( \frac{\lambda+\lambda_0}{\lambda_0 \lambda} \right) \right] 
\nonumber \\
&& \left[-  2 \frac{\lambda_0-\vert X \vert }{\lambda_0}  \theta(0<X<\lambda_0) 
 +  \frac{\lambda +\lambda_0-\vert X \vert }{\lambda+\lambda_0} \theta(0<X<\lambda+\lambda_0)
\right]  +O(\mu^2)
\end{eqnarray}

\subsection{Correction of order $\mu$ to the thermal width }

\subsubsection{Contribution of particles in advance}

The specific contribution of the configurations $M_-M_+M_{++}$
to the thermal width reads with the measure (\ref{measureMMM++})
 \begin{eqnarray}
[c_2]^{(1)}_{M_-M_+M_{++}} && \equiv [c_2]^{(0)+(1)}_{M_-M_+M_{++}}
- [c_2]^{(0)}_{M_-M_+}  = \overline{ (1- e^{-\frac{t}{T_M} }) X_- \lambda }
\label{deltaavance}
 \end{eqnarray}
After the integration over $X_-$ (\ref{firststep}), 
we thus obtain an integral of type (\ref{wpq})
for the values $p=2,q=1$
 \begin{eqnarray}
[c_2]^{(1)}_{M_-M_+M_{++}} = << (1- e^{-\frac{t}{T_M} }) 
\frac{\lambda_0^2}{2} \lambda >>_a
= \frac{1}{2} W_{2,1}(\mu)
 \end{eqnarray}

By symmetry, the configurations $M_{--}M_-M_+$ give exactly the same contribution, and thus the total contribution of particles
in advance reads at lowest order in $\mu$ (\ref{wpqmu0})

 \begin{eqnarray}
[c_2]^{(1)}_{a}
=  W_{2,1}(\mu)
\opsimeq_{\mu \to 0} 
 \mu \int_0^{2 e^{ - \frac{(1+\gamma_E)}{2} }} \frac{dz}{z} 
\left[ 
4 - \frac{z^{5}}{  4 }
 K_{3}(z)
K_{2}(z) \right]
 + O(\mu^2) = \mu \ 0.5383
\label{delta1a}
\end{eqnarray}

\subsubsection{Contribution of particles in delay}

The specific contribution of the configurations $M_-SM_+$
to the thermal width reads with the measure (\ref{measureM-SMM+})
 \begin{eqnarray}
[c_2]^{(1)}_{d} \equiv [c_2]^{(0)+(1)}_{M_-SM_+}
- [c_2]^{(0)}_{M_-M_+}  = - \overline{  e^{-\frac{t}{T_S} }
\left[ \theta(X_S>0) \lambda_+ (\lambda_- -X_S) 
+ \theta(X_S<0) \lambda_- (\lambda_+ +X_S) \right] } 
  \end{eqnarray}
After integration over $X_S$, we thus obtain the integral of type
(\ref{omegapq}) for the special value $(p=2,q=1)$
 \begin{eqnarray}
[c_2]^{(1)}_{d} && =  - <<  e^{-\frac{t}{T_S} }
\left[  \lambda_+ \frac{ \lambda_-^2 }{2} 
+  \lambda_- \frac{ \lambda_-^2 }{2} \right] >>_d
=
-\Omega_{2,1}(\mu)  \nonumber \\
&&  \opsimeq_{\mu \to 0}
- \frac{   \mu }{4}
\int_{ 2 e^{ - \frac{(1+\gamma_E)}{2} }}^{+\infty} dz z^4 K_3(z)
K_2(z) + O(\mu^2) = - \mu 2.2172... +O(\mu^2)
\label{deltaretard}
\end{eqnarray}

\subsubsection{ Total correction of order $\mu$ for the thermal width }

Adding the contributions of particles in advance and in delay, we get
the total correction at order $\mu$
 \begin{eqnarray}
[c_2]^{(1)}_{total}  =  [c_2]^{(1)}_{a} 
+ [c_2]^{(1)}_{d}
 = \mu \left[- \frac{5}{6}  
  - 2   + 2 \gamma_E  
\right] 
 = - \mu \ 1.6789
\label{delta1total}
 \end{eqnarray}

\subsection{Correction of order $\mu$ to the localization parameters }

\subsubsection{Contribution of particles in advance}

For a given configuration $M_-M_+M_{++}$ ,
the localization parameters read (\ref{pMMM++})
 \begin{eqnarray}
[Y_k]^{(0)+(1)}_{M_-M_+M_{++}}
= \left[ e^{-\frac{t}{T_{M}} } \frac{X_-}{\lambda_0} \right]^k
+ \left[ (1-e^{-\frac{t}{T_{M}} } ) \frac{X_-}{\lambda_0+\lambda} \right]^k
+ \left[ 1- e^{-\frac{t}{T_{M}} } \frac{X_-}{\lambda_0}
- (1-e^{-\frac{t}{T_{M}} } ) \frac{X_-}{\lambda_0+\lambda} \right]^k  
 \end{eqnarray}
The average of the specific contribution
may be computed with (\ref{firststep}),
so that taking into account the similar contribution of
the configurations $M_{--}M_-M_+$, we finally obtain
 \begin{eqnarray}
&& \overline{ [Y_k]^{(1)}_{a} }
 = 2 \overline{ \left( [Y_k]^{(0)+(1)}_{M_-M_+M_{++}}
- [Y_k]^{(0)}_{M_-M_+} \right) } \nonumber \\
&& = 2 <<  \frac{\lambda_0}{k+1}
\left[    e^{- k\frac{t}{T_{M}} }   - 1  
+  (1-e^{-\frac{t}{T_{M}} } )^k \frac{ \lambda_0^{k}}
{ (\lambda_0+\lambda)^k} + 
\sum_{m=1}^k (1-e^{-\frac{t}{T_{M}} } )^{m} 
\left( \frac{\lambda}{ \lambda + \lambda_0 } \right)^{m} \right] >>_a
 \end{eqnarray}

In particular, it reads for $k=2$
 \begin{eqnarray}
\overline{ [Y_2]^{(1)}_{a} }
 = << \frac{2 \lambda_0}{3}
\left[ (1-e^{-\frac{t}{T_{M}} } )
\left( \frac{\lambda}{(\lambda + \lambda_0)} -2 \right)
+ (1-e^{-\frac{t}{T_{M}} } )^2 2 
\left( 1- \frac{\lambda \lambda_0}{(\lambda + \lambda_0)^2}  \right)
\right] >>_a
\end{eqnarray}

\subsubsection{Contribution of particles in delay}

For a given configuration $M_-SM_+$ ,
the localization parameters read (\ref{pM-SM+})
for the case $X_S>0$
 \begin{eqnarray}
[Y_k]^{(0)+(1)}_{M_-SM_+} 
= \left[ e^{-\frac{t}{T_{S}} } \frac{\lambda_- - X_S}{\lambda_-} \right]^k
+ \left[ (1-e^{-\frac{t}{T_{S}} } )
 \frac{\lambda_- - X_S}{\lambda_++\lambda_-} \right]^k
+ \left[ 1- e^{-\frac{t}{T_{S}} } \frac{\lambda_- - X_S}{\lambda_-}
- (1-e^{-\frac{t}{T_{S}} } ) \frac{\lambda_- - X_S}{\lambda_++\lambda_-} \right]^k  
 \end{eqnarray}
and a similar expression for $X_S<0$.

The integration over $X_S \in [0,\lambda_-]$ (\ref{firststep2}) yields
 \begin{eqnarray}
 \int_{0}^{\lambda_-} dX_S  [Y_k]^{(0)+(1)}_{M_-SM_+} 
 = \frac{ \lambda_-  }
{ (k+1) } 
\left[ e^{- k \frac{t}{T_{S}} }  
  + (1-e^{-\frac{t}{T_{S}} } )^k
\frac{\lambda_-^{k} }{  (\lambda_+ +\lambda_-)^k}  +  
\sum_{m=0}^k  (1-e^{-\frac{t}{T_{S}} } )^{m} 
\left( \frac{\lambda_+}{ \lambda_+ + \lambda_- } \right)^{m} \right]
 \end{eqnarray}
The case $X_S \in [-\lambda_+,0]$ leads to the symmetric contribution
via the exchange between $(\lambda_+,\lambda_-)$, and thus
the average of the specific
contribution reads
 \begin{eqnarray}
&& \overline {  [Y_k]^{(1)}_{d} }
\equiv << \int_{-\lambda_+}^{\lambda_-} dX_S 
\left( [Y_k]^{(0)+(1)}_{M_-SM_+} - [Y_k]^{(0)}_{M_-M_+}\right) >>_d \\
&& = << \frac{ 1  }
{ (k+1) } 
\left[ (e^{- k \frac{t}{T_{S}} } -1) (\lambda_-+\lambda_+)  
  + (1-e^{-\frac{t}{T_{S}} } )^k 
\frac{\lambda_-^{k+1}+\lambda_+^{k+1} }
{  (\lambda_+ +\lambda_-)^k}  +  
\sum_{m=1}^k  (1-e^{-\frac{t}{T_{S}} } )^{m} 
\left( \frac{\lambda_- \lambda_+^m + \lambda_+ \lambda_-^m}{ (\lambda_+ + \lambda_-)^m } \right) \right] >>_d \nonumber
 \end{eqnarray}

In particular, it reads for $k=2$
 \begin{eqnarray}
\overline {  [Y_2]^{(1)}_{d} }=
- << \frac{2}{3} e^{-\frac{t}{T_{S}} } (1-e^{-\frac{t}{T_{S}} })
\left[ \lambda_- + \lambda_+
- \frac{\lambda_- \lambda_+}{ \lambda_- + \lambda_+ } \right] >>_d
\end{eqnarray}

\subsection{Correction to Aging and Sub-aging properties}

\subsubsection{ Probability $\Pi(t_w+t,t_w)$ of no jump during he interval $[t_w,t_w+t]$}

Since the probability $\Pi(t_w+t,t_w)$ of no jump during he interval $[t_w,t_w+t]$
is directly related to the probability $\psi_{t_w}(\tau)$
 to be at time $t_w$ in a trap of trapping time $\tau$
via (\ref{defpittprime}), the configurations $M_-M_+M_{++}$
do not give any correction, since the trapping time of $M_{++}$
has the same statistical properties as the trapping times of $M_-$ and $M_+$.

On the contrary, the configurations $M_-SM_+$ will give a correction
to $\psi_{t}(\tau)$. For the case $X_S>0$, we have 
 \begin{eqnarray}
[\psi_{t}(\tau)]_{M_-SM_+}^{(0)+(1)}
&&  =  e^{-\frac{t}{T_{S}} } \frac{\lambda_- - X_S}{\lambda_-} \delta(\tau-\tau_S)
+  (1-e^{-\frac{t}{T_{S}} } )
 \frac{\lambda_- - X_S}{\lambda_++\lambda_-} \delta(\tau-\tau_{M_+}) 
\nonumber \\
&& + \left[ 1- e^{-\frac{t}{T_{S}} } \frac{\lambda_- - X_S}{\lambda_-}
- (1-e^{-\frac{t}{T_{S}} } ) \frac{\lambda_- - X_S}{\lambda_++\lambda_-}
\right]
\delta(\tau-\tau_{M_-})   
 \end{eqnarray}
so that the specific contribution yields after integration over $X_S \in [0,\lambda_-]$
 \begin{eqnarray}
&& \int_0^{\lambda_-} dX_S [\psi_t(\tau)]_{M_-SM_+}^{(1)}
 = \int_0^{\lambda_-} dX_S
\left( [\psi_t(\tau)]_{M_-SM_+}^{(0)+(1)}
- [\psi_t(\tau)]_{M_-M_+}^{(0)} \right) \nonumber \\
&&  =  e^{-\frac{t}{T_{S}} } \frac{\lambda_- }{2}
\left[  \delta(\tau-\tau_S) - \delta(\tau-\tau_{M_-}) \right]
  + e^{-\frac{t}{T_{S}} } 
 \frac{\lambda_-^2}{ 2( \lambda_++\lambda_- )}
\left[ \delta(\tau-\tau_{M_-}) - \delta(\tau-\tau_{M_+}) \right]
 \end{eqnarray}
After averaging over $\tau_{M_+}$ and $\tau_{M_-}$, the second term
will vanish. Since the case $X_S<0$ gives the symmetric contribution
via the exchange $(\lambda_+,\lambda_-)$, we finally
obtain that the full correction to the
probability distribution reads with the measure (\ref{meanwithd})
 \begin{eqnarray}
\overline{ [\psi_t(\tau)]^{(1)} }
&& = \int_{R(t)}^{+\infty} \frac{\mu}{\tau_M} \left( \frac{R(t)}{\tau_M} \right)^{\mu} << 
e^{-\frac{t}{T_{S}} } \frac{\lambda_- +\lambda_-}{2}
\left[  \delta(\tau-\tau_S) - \delta(\tau-\tau_{M_-}) \right] >>_d
\end{eqnarray}
We thus finally obtain at first order in $\mu$
 \begin{eqnarray}
\overline{ [\psi_t(\tau)]^{(1)} }
&& = \theta(\tau< R(t)  )  
\frac{\mu}{2 \tau } z_0^3(0)
\left( \frac{R(t)}{\tau} \right)^{\frac{3}{2}}
K_1 \left( z_0(0) \left( \frac{R(t)}{\tau} \right)^{\frac{1}{2}} \right)
K_2 \left( z_0(0) \left( \frac{R(t)}{\tau} \right)^{\frac{1}{2}} \right)
\nonumber \\
&& -  \Omega_{0,1} (\mu) \theta(R(t) < \tau )  
\frac{\mu}{\tau} \left( \frac{R(t)}{\tau} \right)^{\mu}
\end{eqnarray}
with (\ref{z0mu0}). The prefactor $\Omega_{0,1} (\mu)$ (\ref{omegapq})
of the second term represents the probability
 to be in a trap of type $S$ at time $t$ and is of order $\mu$
(\ref{pid}).
So in the domain $\tau>R$, the total distribution $
[\psi_t(\tau)]^{(0)+(1)}$ of $\tau$ keeps
the same form that at zeroth order, but the amplitude 
is $(1-\Omega_{0,1} (\mu))$. There is now
a contribution of the domain $\tau<R$ which was absent at zeroth order.
In particular, in the limit $\tau \to 0$, we obtain the essential singularity
 \begin{eqnarray}
 [\psi_t(\tau)]^{(1)}
\opsimeq_{\tau \to 0} 
\frac{\mu \pi  z_0^2(0) R(t) }{4 \tau^2 }
e^{- 2 z_0(0) \left( \frac{R(t)}{\tau} \right)^{\frac{1}{2}} }
\end{eqnarray}

We now compute the correction to the scaling function ${\tilde \Pi} ( v )$
(\ref{piscalingform}) at first order in $\mu$
\begin{eqnarray}
 {\tilde \Pi}^{(1)} (v)
= \mu 
\int_{z_0(0)}^{+\infty} dz 
z^{2}
K_1 (z)
K_2 (z) e^{-  v \left( \frac{z}{z_0(0)} \right)^2 }
 -  \Omega_{0,1} (\mu) \int_0^1 dy \mu y^{\mu-1}
 e^{- y v }
\end{eqnarray}

So for large $v$, the only correction to the scaling function
${\tilde \Pi}^{(0)} (v)$ (\ref{piscal0})
comes from the second term 
\begin{eqnarray}
 {\tilde \Pi}^{(1)} (v)
\opsimeq_{v \to \infty} 
 -  \Omega_{0,1} (\mu) \frac{ \mu }{v^{\mu} }
\end{eqnarray}
which corresponds to a correction of order $\mu^2$ 
for the amplitude of the algebraic decay $1/v^{\mu}$.

\subsubsection{ 
 Probability $C(t+t_w,t_w)$ to be at time $(t+t_w)$ in the same trap
 at it was at time $t_w$ }

Since in our framework, $C(t+t_w,t_w)$ is determined by the probability
$\phi_{t_w}(T)$ to be at time $t_w$ in a trap of escape time $T$
(\ref{ctophi}), we obtain as for $\Pi(t_w+t,t_w)$
that the configurations $M_-M_+M_{++}$
do not give any correction, because the escape time of $M_{++}$
has the same statistical properties as the escape times of $M_-$ and $M_+$.

The correction due to the configurations $M_-SM_+$ 
can be written for the case $X_S>0$ as 
 \begin{eqnarray}
[\phi_t(T)]_{M_-SM_+}^{(0)+(1)}
&&  =  e^{-\frac{t}{T_{S}} } \frac{\lambda_- - X_S}{\lambda_-} \delta(T-T_S)
+  (1-e^{-\frac{t}{T_{S}} } )
 \frac{\lambda_- - X_S}{\lambda_++\lambda_-} \delta(T-T_{M_+}) 
\nonumber \\
&& + \left[ 1- e^{-\frac{t}{T_{S}} } \frac{\lambda_- - X_S}{\lambda_-}
- (1-e^{-\frac{t}{T_{S}} } ) \frac{\lambda_- - X_S}{\lambda_++\lambda_-}
\right]
\delta(T-T_{M_-})   
 \end{eqnarray}
so that the specific contribution yields after integration over $X_S \in [0,\lambda_-]$
 \begin{eqnarray}
&& \int_0^{\lambda_-} dX_- [\phi_t(\tau)]_{M_-SM_+}^{(1)}
 = \int_0^{\lambda_-} dX_-
\left( [\phi_t(T)]_{M_-SM_+}^{(0)+(1)}
- [\phi_t(\tau)]_{M_-M_+}^{(0)} \right) 
\nonumber \\
&&  =  e^{-\frac{t}{T_{S}} } \frac{\lambda_- }{2}
\left[  \delta(T-T_S) - \delta(T-T_{M_-}) \right]
  + e^{-\frac{t}{T_{S}} } 
 \frac{\lambda_-^2}{ 2( \lambda_++\lambda_- )}
\left[ \delta(T-T_{M_-}) - \delta(T-T_{M_+}) \right]
 \end{eqnarray}
After averaging over $T_{M_+}$ and $T_{M_-}$, the second term
will vanish. Since the case $X_S<0$ gives the symmetric contribution
via the exchange $(\lambda_+,\lambda_-)$, we finally
obtain that the full correction to the
probability distribution reads with the measure (\ref{meanwithd})

 \begin{eqnarray}
\overline [\phi_t(T)]^{(1)}
&&  = << \int_R^{+\infty} \mu \frac{ d \tau_M}{\tau_M} \left(
\frac{R}{\tau_M} \right)^{\mu}
 e^{-\frac{t}{T_{S}} } \frac{\lambda_- +\lambda_+}{2}
\left[  \delta(T-T_S) - \delta(T-T_{M_-}) \right] >>_d 
\nonumber \\
&& = \mu \int_0^R \frac{ d \tau_S}{\tau_S} \left(\frac{R}{\tau_S} \right)^{\mu}
\int_0^{+\infty} d \lambda_+ \int_0^{+\infty} d \lambda_-
 e^{-(\lambda_+ + \lambda_- ) \left( \frac{R(t)}{\tau_S} \right)^{\mu} }
\lambda_-
e^{- \frac{t}{T} }
\delta  
\left( T- \frac{ 2 R^{\mu} \tau_S }{   
\left( \frac{1}{\lambda_+} 
+ \frac{1}{\lambda_-}  \right) } \right)
\nonumber  \\
&& - \mu \int_0^R \frac{ d \tau_S}{\tau_S} \left(
\frac{R}{\tau_S} \right)^{\mu}
\int_0^{+\infty} d \lambda_+ \int_0^{+\infty} d \lambda_-
 e^{-(\lambda_+ + \lambda_- ) \left( \frac{R(t)}{\tau_S} \right)^{\mu} }
\lambda_-
e^{- \frac{t}{2 R^{\mu} \tau_S}
 \left( \frac{1}{\lambda_+} + \frac{1}{\lambda_-}  \right) }
\nonumber \\
&& \int_0^{+\infty} d \lambda e^{- \lambda}
\int_R^{+\infty} \mu \frac{ d \tau_M}{\tau_M} \left(
\frac{R}{\tau_M} \right)^{\mu}
\delta \left( T - \frac{ 2 R^{\mu} \tau_M}{   \left( \frac{1}{\lambda} 
+ \frac{1}{\lambda_++\lambda_-}  \right)}  \right)
\end{eqnarray}
so the correction to the scaling function  ${\tilde C}_{\mu}(h)$ 
(\ref{defcscalw}) reads 
 \begin{eqnarray}
&&  {\tilde C}_{\mu}^{(1)} (h) = \int_0^{+\infty} dT 
\overline [\phi_t(T)]^{(1)}
 e^{- h  \frac{t}{T} } 
\nonumber \\
&& = \mu \int_1^{+\infty} \frac{ d v}{v} v^{\mu}
\int_0^{+\infty} d \lambda_+ \int_0^{+\infty} d \lambda_-
 e^{-(\lambda_+ + \lambda_- ) v^{\mu} }
\lambda_-
e^{- (1+h) \frac{{\tilde T}_0(0)}{2  } v \left( \frac{1}{\lambda_+} 
+ \frac{1}{\lambda_-}  \right) }
 \nonumber \\
&& - \mu \int_1^{+\infty} \frac{ d v}{v} v^{\mu}
\int_0^{+\infty} d \lambda_+ \int_0^{+\infty} d \lambda_-
 e^{-(\lambda_+ + \lambda_- ) v^{\mu} }
\lambda_-
e^{- \frac{{\tilde T}_0(0)}{2  } v
 \left( \frac{1}{\lambda_+} + \frac{1}{\lambda_-}  \right) }
\nonumber \\
&& \int_0^{+\infty} d \lambda e^{- \lambda}
\int_0^{1} \mu \frac{ d w}{w} w^{\mu}
e^{ - h \frac{ {\tilde T}_0(0) }{2  } w \left( \frac{1}{\lambda} 
+ \frac{1}{\lambda_++\lambda_-}  \right)  }
\end{eqnarray}

\section{Set of the important configurations at order $n$}

\label{configordern}

With the $T_n$ traps described in the text (\ref{numbertn}), 
we have to construct the possible 
$\Omega_n$ configurations of
$(2+n)$ traps, that are ordered in positions,
and that contribute up to order $\mu^n$. 
We have
\begin{eqnarray} 
\Omega_n=\Omega_{n-1}+\omega_n = \sum_{i=0}^n \omega_i
\end{eqnarray}
 where $\omega_n$
represents the number of configurations that begin to contribute
at order $n$. 
We may decompose
\begin{eqnarray} 
\omega_n= \sum_{k \geq 1, l\geq 1, k+l \leq 2+n} a_n^{(M_{k-},M_{l+})}
\end{eqnarray}
where $a_n^{(M_{k-},M_{l+})}$ is the number of configurations
that contain $M_{k-}$ as leftmost trap and $M_{l+}$ as rightmost trap.
For $(k=1,l=1+n)$, there is only $a_n^{(M_{-},M_{(1+n)+})}=1$ configuration $\{M_-,M_+,M_{2+},..M_{(n+1)+}\}$,
whereas for $(k=1,l=1)$, there are
\begin{eqnarray} 
a_n^{(M_{-},M_{+})}=n!
\end{eqnarray}
configurations, since we have to order in space the $n$ traps 
$S_1^{(0)}$...$S_n^{(0)}$  
in the interval $]M_-,M_+[$.
More generally, at order $(k,l)$, to construct the configurations 
of $(n+2)$ traps containing $M_{k-} M_{(k-1)-}, .. M_-,M_+,M_{2+},..M_{l+}$,
which represent $(k+l)$ fixed traps,
we have to choose $(n+2-k-l)$ traps among the $(k+l-1)$ available intervals
and to count the possible positional orders in each interval
\begin{eqnarray} 
a_n^{(M_{k-},M_{l+})}=\sum_{p_1=0}^{+\infty} ... \sum_{p_{k+l-1}=0}^{+\infty}
\delta \left(\sum_{i=1}^{k+l-1} p_i=n+2-k-l \right) p_1! ... p_{k+l-1}!
\end{eqnarray}

The final result is thus that the number of new configurations
that appear at order $n$ reads
 \begin{eqnarray} 
\omega_n=\sum_{k \geq 1, l\geq 1, k+l \leq 2+n}
 \left[ \sum_{p_1=0}^{+\infty} ... \sum_{p_{k+l-1}=0}^{+\infty}
\delta(\sum_{i=1}^{k+l-1} p_i=n+2-k-l) p_1! ... p_{k+l-1}!
\right]
\end{eqnarray}
which generalize what we have found before for the lowest orders
with 
$\omega_0=1$ corresponding to $(M_-,M_+)$,
and 
$\omega_1=3$ corresponding to $(M_{--},M_-,M_+)$, $(M_-,M_+,M_{++})$ 
and $(M_-,S,M_+)$.

\section{Useful properties of Bessel functions}

\label{secbessel}

The following integrals yields Bessel function of type $K$
 \begin{eqnarray}
\int_0^{+\infty} dx x^{\nu} e^{-a x -\frac{b}{x} }
= 2 \left( \frac{b}{a}  \right)^{\frac{1+\nu}{2} } K_{1+\nu}(2 \sqrt {a b}) 
\label{inteleadingtoK}
 \end{eqnarray}

The asymptotic behavior at infinity is independent of $\nu$ and reads
 \begin{eqnarray}
 K_{\nu}(z) \opsimeq_{z \to \infty} \sqrt{ \frac{\pi}{2 z} } e^{-z}
\label{Kzgrand}
 \end{eqnarray}
Near the origin, the behavior depends on $\nu$.
We will need the behavior for $\nu=1$
 \begin{eqnarray}
 K_{1}(z) \opsimeq_{z \to 0} \frac{1}{z} + \frac{z}{2} 
\left[ \ln \frac{z}{2}+\gamma_{Euler} - \frac{1}{2}   \right] + 
O(z^3 \ln z)
\label{K1zpetit}
 \end{eqnarray}

Another useful integral is
 \begin{eqnarray}
\int_0^{+\infty} dz z^{2k-1} K_{\nu}^2(z)
= \frac{{ \sqrt \pi} \Gamma(k) \Gamma(k+\nu) \Gamma(k-\nu)}{ 4 \Gamma \left(k+\frac{1}{2} \right) }
\label{inteK2power}
 \end{eqnarray}

\end{document}